\def\Kr{\kA\bds\cdot\uR}
\def\Kl{\kA\bds\cdot\ul}
\def\Km{\kA\bds\cdot\um}

\def\Kn{\kA\bds\cdot\uN}

\def\ul{{\bds{\hat{l}}}}
\def\um{{\bds{\hat{m}}}}

\def\sA{{\bds{\mathcal{S}}}^\textrm{A}}
\def\sB{{\bds{\mathcal{S}}}^\textrm{B}}

\def\kA{{\bds{\hat{S}}}^\textrm{A}}
\def\kB{{\bds{\hat{S}}}^\textrm{B}}

\def\mA{M_\textrm{A}}
\def\mB{M_\textrm{B}}
\def\uR{\boldsymbol{\hat{\rho}}}
\def\uT{\boldsymbol{\hat{\tau}}}
\def\uN{\boldsymbol{\hat{\nu}}}
\def\uI{{\boldsymbol{\hat{e}}}_x}
\def\uJ{{\boldsymbol{\hat{e}}}_y}
\def\uK{{\boldsymbol{\hat{e}}}_z}
\def\nk{n_{\rm b}}

\def\snf{\sin f}
\def\csf{\cos f}
\def\cu{\cos u}
\def\su{\sin u}

\def\Pb{P_{\rm b}}

\def\rfr#1{Equation~(\ref{#1})}
\def\rfrs#1#2{Equations~(\ref{#1})~to~(\ref{#2})}
\def\Rfr#1{Eq. (\ref{#1})}

\def\derp#1#2{\rp{\partial{#1}}{\partial{#2}}}
\def\dert#1#2{\frac{{{\textrm{d}}}{#1}}{{{\textrm{d}}}{#2}}}

\def\virg#1{``#1"}

\def\eqi{\begin{equation}}
\def\eqf{\end{equation}}
\def\eqia{\begin{eqnarray}}
\def\eqfa{\end{eqnarray}}

\def\rp#1#2{{#1\over#2}}
\def\lb#1{\label{#1}}

\def\bds#1{\boldsymbol{#1}}

\def\co{\cos\omega}
\def\so{\sin\omega}

\def\cO{\cos\Omega}
\def\sO{\sin\Omega}

\def\cI{\cos I}
\def\sI{\sin I}

\def\ton#1{\left(#1\right)}
\def\qua#1{\left[#1\right]}
\def\grf#1{\left\{#1\right\}}
\def\ang#1{\left\langle #1\right\rangle}
\documentclass{aastex}

\usepackage{hyperref}
\usepackage{float}
\usepackage{amsmath,textgreek,w-greek,wasysym}
\usepackage{amscd,lineno}
\usepackage{amssymb}
\usepackage{graphicx,epsfig}
\usepackage{txfonts}
\usepackage{xr-hyper}

\RequirePackage{color}

\newcommand{\emaila}{lorenzo.iorio@libero.it}

\allowdisplaybreaks[1]

\begin{document}

\title{Post-Keplerian effects on radial velocity in binary systems and the possibility of measuring General Relativity with the star S2 in 2018}

\shortauthors{L. Iorio}

\author{Lorenzo Iorio\altaffilmark{1} }
\affil{Ministero dell'Istruzione, dell'Universit\`{a} e della Ricerca
(M.I.U.R.)-Istruzione
\\ Permanent address for correspondence: Viale Unit\`{a} di Italia 68, 70125, Bari (BA),
Italy}

\email{\emaila}

\begin{abstract}
One of the directly measured quantities used in monitoring the orbits of many of the S stars revolving around the Supermassive Black Hole (SMBH) in the Galactic Center (GC) is their radial velocity (RV) $V$ obtained with near-infrared spectroscopy. Here, we devise a general approach to calculate both the instantaneous variations $\Delta V\ton{t}$ and the net shifts  per revolution $\ang{\Delta V}$ induced on such an observable by some post-Keplerian (pK) accelerations. In particular, we look at the general relativistic Schwarzschild (gravitoelectric) and Lense-Thirring (gravitomagnetic frame-dragging) effects, and the mass quadrupole. It turns out that we may be on the verge of measuring the Schwarzschild-type 1pN static component of the SMBH's field with the star S2 for which RV measurements accurate to about $\simeq 30-50~\textrm{km~s}^{-1}$ dating back to $t_0 = 2003.271$ are currently available, and whose orbital period amounts to $\Pb = 16$ yr. Indeed, while its expected general relativistic RV net shift per orbit amounts to just $\ang{\Delta V^\textrm{GE}} = -11.6~\textrm{km~s}^{-1}$, it should reach a peak value as large as $\Delta V_\textrm{max}^\textrm{GE}\ton{t_\textrm{max}} = 551~\textrm{km ~s}^{-1}$ at $t_\textrm{max} = 2018.35$. \textcolor{black}{Current uncertainties in the S2 and SMBH's estimated parameters yield a variation range from $505~\textrm{km~s}^{-1}~(2018.79)$ to $575~\textrm{km~s}^{-1}~(2018.45)$. }
The periastron shift $\Delta\omega^\textrm{GE}$ of S2 over the same time span will not be larger than $0.2$ deg, while the current accuracy in estimating such an orbital element for this star is of the order of $0.6$ deg.
The frame-dragging and quadrupole-induced RV shifts are far smaller for S2, amounting to, at most, $0.19~\textrm{km~s}^{-1},0.0039~\textrm{km~s}^{-1}$, respectively.
\end{abstract}

keywords{
gravitation--celestial mechanics--binaries: general--pulsars: general--stars: black holes
}
\section{Introduction}
Let us consider a gravitationally bound binary system made by two extended rotating bodies A and B orbiting about their common center of mass. Let B denote the invisible companion of the binary's component A for which a light curve is, instead, spectroscopically available through Doppler measurements of its radial velocity (RV) at some wavelengths of the electromagnetic spectrum. In an exoplanetary system, the body A is typically the host star while B is the directly undetectable planet. In a binary hosting a main sequence star and a compact object like a white dwarf, a neutron star or a Black Hole (BH), A is usually the star and B is the unseen stellar corpse. In the stellar system surrounding the Supermassive Black Hole (SMBH) in the Galactic Center at Sgr A$^\ast$ \citep{2017ApJ...837...30G}, A is some of the revolving S-type stars, while B is the SMBH itself. By assuming a coordinate system centered in the binary's center of mass whose reference $z$ axis points toward the observer along the line of sight in such a way that the reference $\grf{x,~y}$ plane coincides with the plane of the sky, the portion of the binary's orbital motion which is directly accessible to observation through spectroscopic measurements is the projection of the barycentric orbit ${\mathbf{r}}_\textrm{A}$ of A onto the line of sight, i.e., to the Keplerian level, \eqi \mathrm{r}_z^\textrm{A} = \rp{\mB}{M_\textrm{tot}}\mathrm{r}_z,\eqf
where\footnote{Appendix \ref{appen} should be consulted for notations and definitions used throughout the text. }
\eqi
\mathrm{r}_z = r\sI\sin u = \rp{a\ton{1-e^2}\sI\sin u}{1 + e\cos f}\lb{zKep}
\eqf
in which we used
\eqi
r\ton{f} = \rp{a\ton{1-e^2}}{1+e\cos f}.\lb{Kepless}
\eqf
The binary's RV $V$, i.e. the projection of the binary's velocity vector $\mathbf{v}$ onto the line of sight, can be straightforwardly obtained by taking the time derivative of \rfr{zKep} performed by using the Keplerian expression
\eqi
\dert f t = \rp{\sqrt{\mu p}}{r^2}= \rp{\nk\ton{1+e\cos f}^2}{\ton{1-e^2}^{3/2}}.\lb{dfdtKep}
\eqf
Thus, the RV of the body A which is provided by its light curve inferred spectroscopically in some portion of the electromagnetic spectrum is, up to the RV $V_0$ of the binary's center of mass,
\eqi
V_\textrm{A} =\rp{\nk a_\textrm{A} \sI}{\sqrt{1-e^2}}\ton{e\co + \cu}=\rp{\nk\mB a \sI}{M_\textrm{tot}\sqrt{1-e^2}}\ton{e\co + \cu}.\lb{RV}
\eqf

In this paper, we illustrate a relatively simple and straightforward approach to analytically calculate the impact that several post-Keplerian (pK) features of motion, both Newtonian and post-Newtonian (pN), have on such a key observable. \textcolor{black}{To the best of our knowledge, there are no analogous analytical calculations in the literature}. Our strategy has a general validity since, in principle, it can be extended to a wide range of dynamical effects, irrespectively of their physical origin, which may  include, e.g., alternative models of gravity as well; think about, e.g., very wide, slowly revolving binaries and long-range modified models of the gravitational interaction. Furthermore, it is applicable to systems whose constituents may have arbitrary masses and compositions (exoplanets, ordinary binary stars, binaries hosting detectable compact objects, transmitting spacecraft orbiting planets or satellites, etc.) and orbital configurations.
Thus, more realistic sensitivity analyses, aimed to both re-interpreting already performed studies and designing future targeted ones, could be conducted in view of a closer correspondence with which is actually measured.
In particular, here, we will focus on some known pK accelerations, both Newtonian (quadrupole) and pN to the lowest order (1pN static and stationary fields).
In principle, our results may be applicable even to anthropogenic binaries like, e.g., those contrived in past concept studies to perform tests of fundamental physics in space \citep{1988AJ.....95..576N,2008IJMPD..17..453S}, or continuously emitting transponders placed on the surface of some moons of larger astronomical bodies.

As a direct application of our results, we will look at the star S2 in Sgr A$^\ast$: the relevant physical and orbital parameters of such a system are listed in Table \ref{S2}. Indeed, it will reach the periastron next year, thus providing a good opportunity to detect the Schwarzschild-type 1pN static field of the Galactic SMBH through its light curve measured in the near-infrared. To this aim, it is important to remark that the currently available RV data records for most of the S stars in Sgr A$^\ast$ cover just more or less extended fractions of their orbital periods which are many years long. Thus, the RV net shifts per orbit $\ang{\Delta V}$ will necessarily play a minor role because of their scarce availability, even in the near-mid future. Suffice it to say that, according to table 3 of \citet{2017ApJ...837...30G}, the fastest S stars for which accurate RV data are available since $t_0=2003.271$ are S2 ($\Pb = \textrm{16~yr}$) and S38 ($\Pb = \textrm{19.2~yr}$). This implies that also the instantaneous, or short-period in the language of celestial mechanics, shifts $\Delta V\ton{t}$ and their temporal patterns are important in view of possible detection of pN effects of interest. Our work will deal also with them. \textcolor{black}{It must be stressed that, in addition to the orbital component treated here, the spectroscopic measurements of S2 account also for several other effects pertaining the propagation of the electromagnetic waves through the deformed spacetime of the SMBH like, e.g., the R{\o}mer time delay, the transverse Doppler shift, the gravitational redshift, the gravitational lensing. A complete investigation of the actual measurability of non-Newtonian gravity in the S2-Sgr A$^\ast$ system may not leave them out, and dedicated analyses, which are outside the scopes of the present paper, should be performed. Examples of them can be found, e.g., in \citet{2006ApJ...639L..21Z,2010ApJ...711..157A,2015ApJ...809..127Z,2016ApJ...827..114Y,2017ApJ...834..198Z}, although they generally deal with temporal windows different from ours.  }

The paper is organized as follows. Section \ref{metodo} details the calculational approach. The 1pN Schwarzschild-type gravitoelectric effects are calculated in Section \ref{GEeffects}, while Section \ref{LTeffects} is devoted to the 1pN gravitomagnetic ones.  The impact of the quadrupole mass moments of the binary's components are treated in Section \ref{Qeffects}.  Section \ref{fine} summarizes our findings.
\section{Outline of the proposed method}\lb{metodo}
If the motion of the binary is affected by some relatively small pK acceleration $\bds A$, either Newtonian or pN in nature,  its impact on the RV of the visible component A  can be calculated perturbatively as follows.
 \citet{1993CeMDA..55..209C} analytically worked out the instantaneous changes $\Delta\mathrm{v}_{\rho},~\Delta\mathrm{v}_{\tau},~\Delta\mathrm{v}_{\nu}$  of the radial, transverse and out-of-plane components $\mathrm{v}_{\rho},~\mathrm{v}_{\tau},~\mathrm{v}_{\nu}$ of the velocity vector $\mathbf{v}$ of the relative motion of a test particle about its primary: they are
\begin{align}
\Delta \mathrm{v}_{\rho}\ton{f} \lb{vR} \nonumber & = -\rp{\nk a \sin f}{\sqrt{1-e^2}}\qua{\rp{e}{2a}\Delta a\ton{f} + \rp{a}{r\ton{f}}\Delta e\ton{f} } -\rp{\nk a^3}{r^2\ton{f}}\Delta{\mathcal{M}}\ton{f}  - \\ \nonumber \\
&-\rp{\nk a^2}{r\ton{f}}\sqrt{1-e^2}\qua{\cI\Delta\Omega\ton{f} + \Delta\omega\ton{f}}, \\ \nonumber \\
\Delta \mathrm{v}_{\tau}\ton{f} \lb{vT} & = -\rp{\nk a\sqrt{1-e^2}}{2r\ton{f}}\Delta a\ton{f} + \rp{\nk a \ton{e+\cos f} }{\ton{1-e^2}^{3/2}}\Delta e\ton{f} +\rp{\nk a e \sin f}{\sqrt{1-e^2}}\qua{\cI\Delta\Omega\ton{f} + \Delta\omega\ton{f}}, \\ \nonumber \\
\Delta \mathrm{v}_{\nu}\ton{f} \lb{vN} & = \rp{\nk a}{\sqrt{1-e^2}}\qua{\ton{\cos u + e \cos\omega}\Delta I\ton{f} + \ton{\sin u + e \so}\sin I\Delta\Omega\ton{f}  }.
\end{align}
In \rfrs{vR}{vN}, the instantaneous changes $\Delta a\ton{f},~\Delta e\ton{f},~\Delta I\ton{f},~\Delta\Omega\ton{f},~\Delta\omega\ton{f}$  are to be calculated as
\eqi
\Delta\kappa\ton{f}=\int_{f_0}^f\dert{\kappa}{t} \dert{t}{f^{'}} df^{'},~\kappa=a,~e,~I,~\Omega,~\omega,\lb{Dk}
\eqf
where the time derivatives $d\kappa/dt$ of the Keplerian orbital elements $\kappa$ are to be taken from the right-hand-sides of the Gauss equations
\begin{align}
\dert a t \lb{dadt}& = \rp{2}{\nk\sqrt{1-e^2}}\qua{e A_{\rho} \snf + A_{\tau}\ton{\rp{p}{r}}}, \\ \nonumber \\
\dert e t \lb{dedt}& = \rp{\sqrt{1-e^2}}{\nk a}\grf{A_{\rho} \snf + A_{\tau}\qua{\csf + \rp{1}{e}\ton{1 - \rp{r}{a}} } }, \\ \nonumber \\
\dert I t \lb{dIdt}& = \rp{1}{\nk a \sqrt{1 - e^2}}A_{\nu}\ton{\rp{r}{a}}\cu, \\ \nonumber \\
\dert \Omega t \lb{dOdt}& = \rp{1}{\nk a \sI\sqrt{1 - e^2}}A_{\nu}\ton{\rp{r}{a}}\su, \\ \nonumber \\
\dert \omega t \lb{dodt}& = -\cI\dert\Omega t + \rp{\sqrt{1-e^2}}{\nk a e}\qua{ -A_{\rho}\csf + A_{\tau}\ton{1 + \rp{r}{p}}\snf },
\end{align}
evaluated onto the  Keplerian ellipse given by \rfr{Kepless} and assumed as unperturbed reference trajectory; the same holds also for $dt/df$ entering \rfr{Dk}, for which the reciprocal of \rfr{dfdtKep} has to be used. The case of the mean anomaly $\mathcal{M}$ is subtler, and requires more care.
Indeed, in the most general case encompassing the possibility that the mean motion $\nk$ is time-dependent because of some physical phenomena, it can be written as\footnote{The mean anomaly at epoch is denoted as $\eta$ by \citet{Nobilibook87}, $l_0$ by \citet{1991ercm.book.....B}, and $\epsilon^{'}$ by \citet{2003ASSL..293.....B}. It is a \virg{slow} variable in the sense that its time derivative vanishes in the limit $\bds A\rightarrow 0$; cfr. with \rfr{detadt}. } \citep{Nobilibook87,1991ercm.book.....B,2003ASSL..293.....B}
\eqi
\mathcal{M}\ton{t} = \eta + \int_{t_0}^{t} \nk\ton{t^{'}}dt^{'};\lb{Mt}
\eqf
the Gauss equation for the variation of the mean anomaly at epoch is
\citep{Nobilibook87,1991ercm.book.....B,2003ASSL..293.....B}
\eqi
\dert\eta t = - \rp{2}{\nk a}A_{\rho}\ton{\rp{r}{a}} -\rp{\ton{1-e^2}}{\nk a e}\qua{ -A_{\rho}\csf + A_{\tau}\ton{1 + \rp{r}{p}}\snf }\lb{detadt}.
\eqf
If $\nk$ is constant, as in the Keplerian case, \rfr{Mt} reduces to the usual form
\eqi
\mathcal{M}\ton{t}= \eta + \nk\ton{t-t_0}.
\eqf
In general, when a disturbing acceleration is present, the semimajor axis $a$ does vary according to \rfr{dadt}; thus, also the mean motion $\nk$ experiences a change\footnote{We neglect the case $\mu\ton{t}$.}
\eqi
\nk\rightarrow \nk+\Delta\nk\ton{t}
\eqf
which can be calculated in terms of the true anomaly $f$ as
\eqi
\Delta \nk\ton{f}=\derp{\nk}{a}\Delta a\ton{f}= -\rp{3}{2}\rp{\nk}{a}\int_{f_0}^f\dert a t \dert{t}{f^{'}}df^{'}\lb{Dn}
\eqf
by means of \rfr{dadt} and the reciprocal of \rfr{dfdtKep}.
Depending on the specific perturbation at hand, \rfr{Dn} does not generally vanish.
Thus, the total change experienced by the mean anomaly $\mathcal{M}$ due to the disturbing acceleration $\bds A$ can be obtained as
\eqi
\Delta\mathcal{M}\ton{f} = \Delta\eta\ton{f} + \int_{t_0}^{t}\Delta\nk\ton{t^{'}} dt^{'},\lb{anom}
\eqf
where
\begin{align}
\Delta\eta\ton{f} &= \int_{f_0}^f\dert\eta t \dert{t}{f^{'}} df^{'}, \\ \nonumber \\
\int_{t_0}^{t}\Delta\nk\ton{t^{'}} dt^{'} \lb{inte}& = -\rp{3}{2}\rp{\nk}{a}\int_{f_0}^f\Delta a\ton{f^{'}}\dert{t}{f^{'}}df^{'}.
\end{align}
In the literature, the contribution due to \rfr{inte} has been often neglected.
An alternative way to compute the perturbation of the mean anomaly with respect to \rfr{anom} implies the use of the mean longitude $\lambda$ and the longitude of pericenter $\varpi$.
It turns out that\footnote{The mean longitude at epoch is denoted as $\epsilon$ by \citet{Nobilibook87,1989racm.book.....S,1991ercm.book.....B,2003ASSL..293.....B}. It is better suited than $\eta$ at small inclinations \citep{2003ASSL..293.....B}.} \citep{1989racm.book.....S}
\eqi
\Delta\mathcal{M}\ton{f} = \Delta\epsilon\ton{f} - \Delta\varpi\ton{f} + \int_{t_0}^{t}\Delta\nk\ton{t^{'}} dt^{'},\lb{longi}
\eqf
where the Gauss equations for the variation of $\varpi,~\epsilon$ are \citep{Nobilibook87,1989racm.book.....S,1991ercm.book.....B,2003ASSL..293.....B}
\begin{align}
\dert\varpi t & =2\sin^2\ton{\rp{I}{2}}\dert\Omega t +\rp{\sqrt{1-e^2}}{\nk a e}\qua{-A_{\rho}\cos f+A_{\tau}\ton{1+\rp{r}{p}}\sin f}, \\ \nonumber \\
\dert\epsilon t & = \rp{e^2}{1+\sqrt{1-e^2}}\dert\varpi t +2\sqrt{1-e^2}\dert\Omega t-\rp{2}{\nk a}A_{\rho}\ton{\rp{r}{a}}.
\end{align}
It must be remarked that, depending on the specific perturbing acceleration $\bds A$ at hand, the calculation of \rfr{inte} may turn out to be rather uncomfortable.

The instantaneous change experienced by the RV of the binary's relative motion can be extracted from \rfrs{vR}{vN} by taking the  $z$ component $\Delta \mathrm{v}_z$ of the vector
\eqi
\Delta \mathbf{v} = \Delta \mathrm{v}_{\rho}~\uR + \Delta \mathrm{v}_{\tau}~\uT + \Delta \mathrm{v}_{\nu}~\uN
\eqf
expressing the perturbation experienced by the binary's relative velocity vector $\mathbf{v}$.
It is
\begin{align}
\Delta \mathrm{v}_z\ton{f}\nonumber \lb{Dvzf}& = -\rp{\nk\sI\ton{e\co + \cu}}{2\sqrt{1-e^2}}\Delta a\ton{f} + \\ \nonumber \\
\nonumber &+ \rp{\nk a\sI\grf{ 4\cos\ton{2f +\omega} + e\qua{-\cos\ton{f-\omega}  +  4\cu +\cos\ton{3f + \omega}  }  }}{4\ton{1-e^2}^{3/2}}\Delta e\ton{f} + \\ \nonumber \\
\nonumber &+ \rp{\nk a\cI\ton{e\co+\cu}}{\sqrt{1-e^2}}\Delta I\ton{f} - \rp{\nk a\sI\ton{e\so + \su}}{\sqrt{1-e^2}}\Delta \omega\ton{f} - \\ \nonumber \\
& - \rp{\nk a\ton{1+e\csf}^2\sI\su }{\ton{1 - e^2}^2}\Delta\mathcal{M}\ton{f}.
\end{align}
It is possible to express the true anomaly as a function of time through the mean anomaly according to \citet[p.~77]{1961mcm..book.....B}
\eqi
f\ton{t} = \mathcal{M}\ton{t} + 2\sum_{s = 1}^{s_\textrm{max}}\rp{1}{s}\grf{ J_s\ton{se} + \sum_{j = 1}^{j_\textrm{max}}\rp{\ton{1-\sqrt{1-e^2}}^j}{e^j}\qua{ J_{s-j}\ton{se} + J_{s+j}\ton{se}  }  }\sin s\mathcal{M}\ton{t}, \lb{fMt}
\eqf
where $J_k\ton{se}$ is the Bessel function of the first kind of order $k$ and $s_\textrm{max},~j_\textrm{max}$ are some values of the summation indexes $s,~j$ adequate for the desired accuracy level.
Having at disposal such analytical time series yielding the time-dependent pattern  of \rfr{Dvzf} allows one to easily study some key features of it such as, e.g., its extrema along with the corresponding epochs and the values of some unknown parameters which may enter the disturbing acceleration.
The net change per orbit $\ang{\Delta \mathrm{v}_z}$ can be obtained by calculating \rfr{Dvzf}
with $f=f_0+2\uppi$, and using \rfr{Dk} and \rfrs{anom}{inte} integrated from $f_0$ to $f_0+2\uppi$.

In order to have the RV shifts of the binary's visible component A, \rfr{Dvzf} and its orbit averaged expression have  to be scaled by $\mB M_\textrm{tot}^{-1}$.
%
%

In the following, we will look at three pK dynamical effects: the Newtonian deviation from spherical symmetry of the binary's bodies due to their quadrupole mass moments $J_2^\textrm{A/B}$, and the velocity-dependent 1pN static (gravitoelectric, GE) and stationary (gravitomagnetic) accelerations responsible of the time-honored anomalous Mercury's perihelion precession and the Lense-Thirring (LT) frame-dragging, respectively.
\section{The 1pN gravitoelectric effect}\lb{GEeffects}
Let us start with the static component of the 1pN field which, in the case of our Solar System, yields the formerly anomalous perihelion precession of Mercury  of $\dot\varpi_{\mercury}=42.98~\textrm{arcsec~cty}^{-1}$ \citep{1986Natur.320...39N}.

The 1pN GE, Schwarzschild-type, acceleration of the relative motion is, in General Relativity, \citep{1989racm.book.....S}
\eqi
{\bds A}_\textrm{GE}=\rp{\mu}{c^2 r^2}\grf{\qua{\ton{4 + 2 \xi}\rp{\mu}{r} -\ton{1 + 3\xi}\ton{\mathbf{v}\bds\cdot\mathbf{v}}  + \rp{3}{2}\xi\ton{\mathbf{v}\bds\cdot\uR}^2  }\uR +\ton{4 - 2\xi}\ton{\mathbf{v}\bds\cdot\uR}\mathbf{v} }.\lb{AGE}
\eqf
By projecting \rfr{AGE} onto the radial, transverse, out-of-plane unit vectors $\uR,~\uT,~\uN$, its corresponding components are
\begin{align}
A_{\rho}^\textrm{GE} \lb{ARGE}& =\rp{\mu^2\ton{1 + e \csf}^2 \qua{ \ton{4 - 13\xi} e^2  + 4\ton{3 - \xi} + 8\ton{1 - 2\xi}e\csf  -\ton{8 - \xi}e^2\cos 2f   } }{4 c^2 a^3\ton{1-e^2}^3 }, \\ \nonumber \\
A_{\tau}^\textrm{GE} \lb{ATGE}& = \rp{2\mu^2 \ton{1 + e \csf}^3 \ton{2 - \xi}e\snf}{c^2 a^3\ton{1 - e^2}^3}, \\ \nonumber \\
A_{\nu}^\textrm{GE} \lb{ANGE}& = 0.
\end{align}
Here, we use the true anomaly $f$ since it turns out computationally more convenient.

The resulting net shifts per orbit of the osculating Keplerian orbital elements, obtained by integrating \rfr{Dk} and \rfrs{anom}{inte} from $f_0$ to $f_0+2\uppi$, are
\begin{align}
\ang{\Delta a^\textrm{GE}} \lb{DaeIOGE}&=\ang{\Delta e^\textrm{GE}} = \ang{\Delta I^\textrm{GE}} = \ang{\Delta\Omega^\textrm{GE}} = 0, \\ \nonumber \\
\ang{\Delta\omega^\textrm{GE}}\lb{Domega}& = \ang{\Delta\varpi^\textrm{GE}} = \rp{6\uppi\mu}{c^2 p}, \\ \nonumber \\
\ang{\Delta\mathcal{M}^\textrm{GE}} \lb{DMGE} \nonumber & = \rp{\uppi\mu}{4c^2 a\ton{1-e^2}^2}\grf{8\ton{-9+2\xi}  +4e^4\ton{-6+7\xi} +e^2\ton{-84+76\xi} + \right.\\ \nonumber\\
\nonumber &+\left.3e\qua{ 8\ton{-7+3\xi}  +e^2\ton{-24+31\xi }   }\cos f_0 + \right. \\ \nonumber \\
&+\left. 3e^2\qua{ 4\ton{-5+4\xi}\cos 2f_0  +e\xi\cos 3 f_0  }}.
\end{align}
If, on the one hand, \rfr{Domega} is the well known relativistic pericenter advance per orbit, on the other hand, \rfr{DMGE} represents a novel result which amends several incorrect expressions existing in the literature \citep{1977CeMec..15...21R,2005A&A...433..385I, 2007Ap&SS.312..331I}, mainly because based only on \rfr{detadt}.
Indeed, it turns out that \rfr{inte}, integrated over an orbital revolution, does not vanish.
By numerically calculating \rfr{DMGE} with the physical and orbital parameters of some binary, it can be shown that it agrees with the expression obtainable for $\ang{\Delta\mathcal{M}}$ from Equations~(A2.78e)~to~(A2.78f) by \citet[p.~178]{1989racm.book.....S} in which all the three anomalies $f,~E,~\mathcal{M}$ appear.
It should be remarked that \rfr{DMGE} is an exact result in the sense that no a-priori assumptions on $e$ were assumed. It can be shown that, to the zero order in $e$, \rfr{DMGE} is independent of $f_0$.

We will not explicitly display here the analytical expressions for the instantaneous changes $\Delta\kappa^\textrm{GE}\ton{f},\kappa=a,~e,~I,~\Omega,~\omega,~\Delta\mathcal{M}^\textrm{GE}\ton{f}$ because of their cumbersomeness, especially as far as the mean anomaly is concerned. However, $\Delta\kappa^\textrm{GE}\ton{f},~\kappa=a,~e,~I,~\Omega,~\omega$ can be found in Equations~(A2.78b)~to~(A2.78d) of \citet[p.~178]{1989racm.book.....S}. Equations~(A2.78e)~to~(A2.78f) of \citet[p.~178]{1989racm.book.....S} allow to obtain the instantaneous shift of the mean anomaly, although in terms of the three anomalies $f,~E,~\mathcal{M}$; instead, our (lengthy) expression contains only the true anomaly $f$. See also Equations~(3.1.102)~to~(3.1.107) of \citet[p.~93]{1991ercm.book.....B}.

The net change per revolution of the radial velocity of the spectroscopically detectable binary's component A can be calculated with \rfr{Dvzf}
together with \rfrs{DaeIOGE}{DMGE}, by obtaining
\begin{align}
-\rp{4c^2\ton{1-e^2}^4}{\uppi G\mB\nk\sI}\ang{\Delta V_\textrm{A}^\textrm{GE}} \nonumber \lb{dvzGE}& = \ton{1+e\cos f_0}^2\grf{8\ton{-9+2\xi}  +4e^4\ton{-6+7\xi} +e^2\ton{-84+76\xi} + \right.\\ \nonumber\\
\nonumber &+\left.3e\qua{ 8\ton{-7+3\xi}  +e^2\ton{-24+31\xi }   }\cos f_0 + \right.\\ \nonumber \\
\nonumber &+\left. 3e^2\qua{ 4\ton{-5+4\xi}\cos 2f_0  +e\xi\cos 3 f_0  }}\sin u_0 + \\ \nonumber \\
& + 24\ton{1-e^2}^{5/2}\ton{e\so + \sin u_0}.
\end{align}
It is important to note that, in general, \rfr{dvzGE} is a periodic function of $f_0$; it holds also to the zeroth order in $e$.
The explicit expression of $\Delta V^\textrm{GE}_\textrm{A}\ton{f}$ will not be displayed explicitly here since it is too unwieldy.

\begin{figure*}
\centerline{
\vbox{
\begin{tabular}{cc}
\epsfysize= 5.00 cm\epsfbox{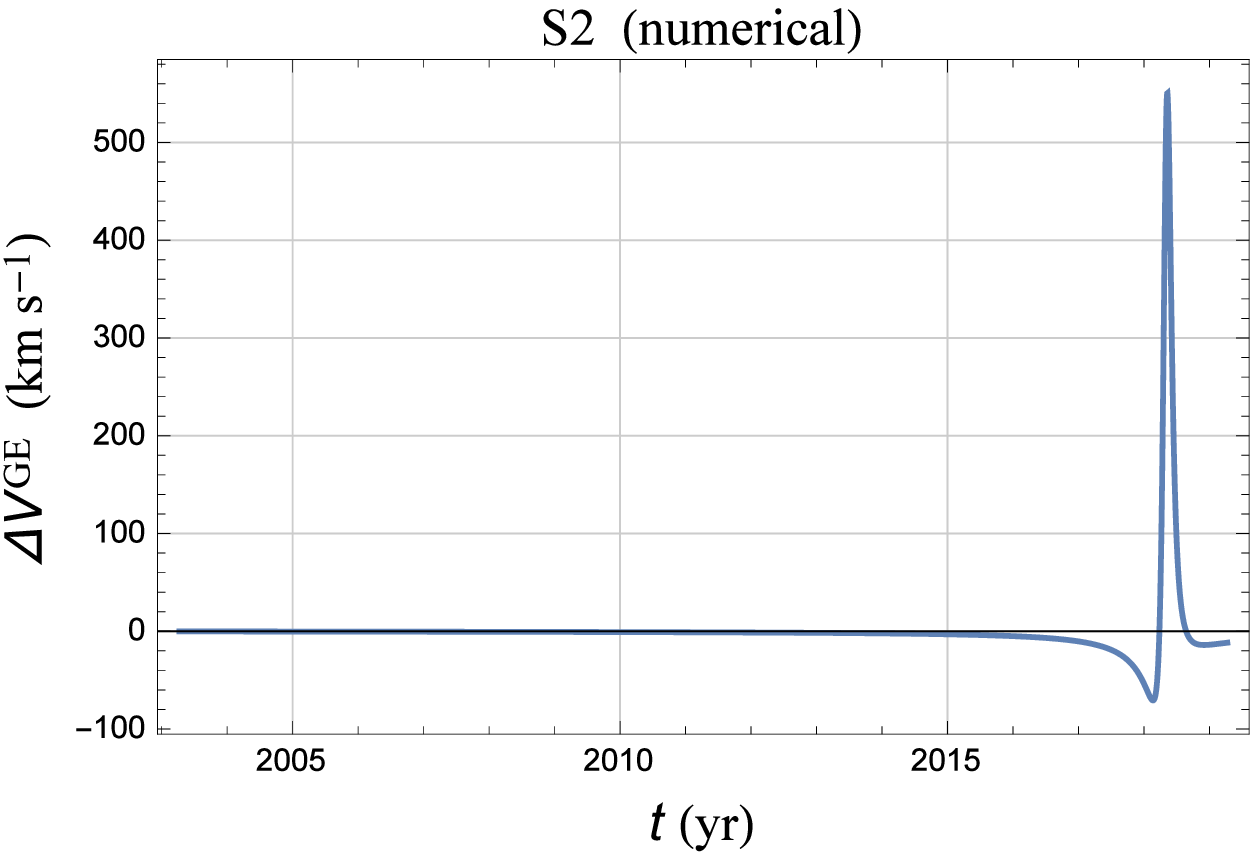}&\epsfysize= 5.00 cm\epsfbox{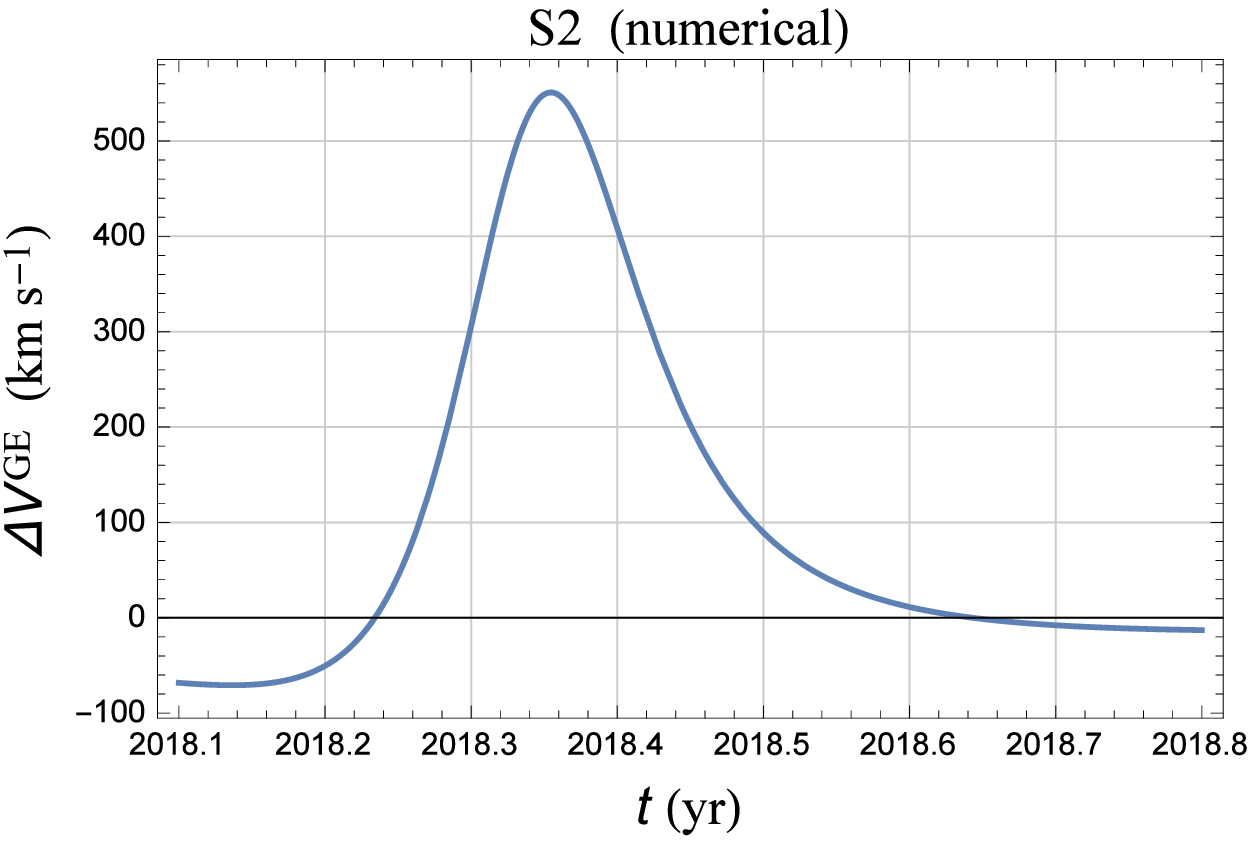}\\
\epsfysize= 5.00 cm\epsfbox{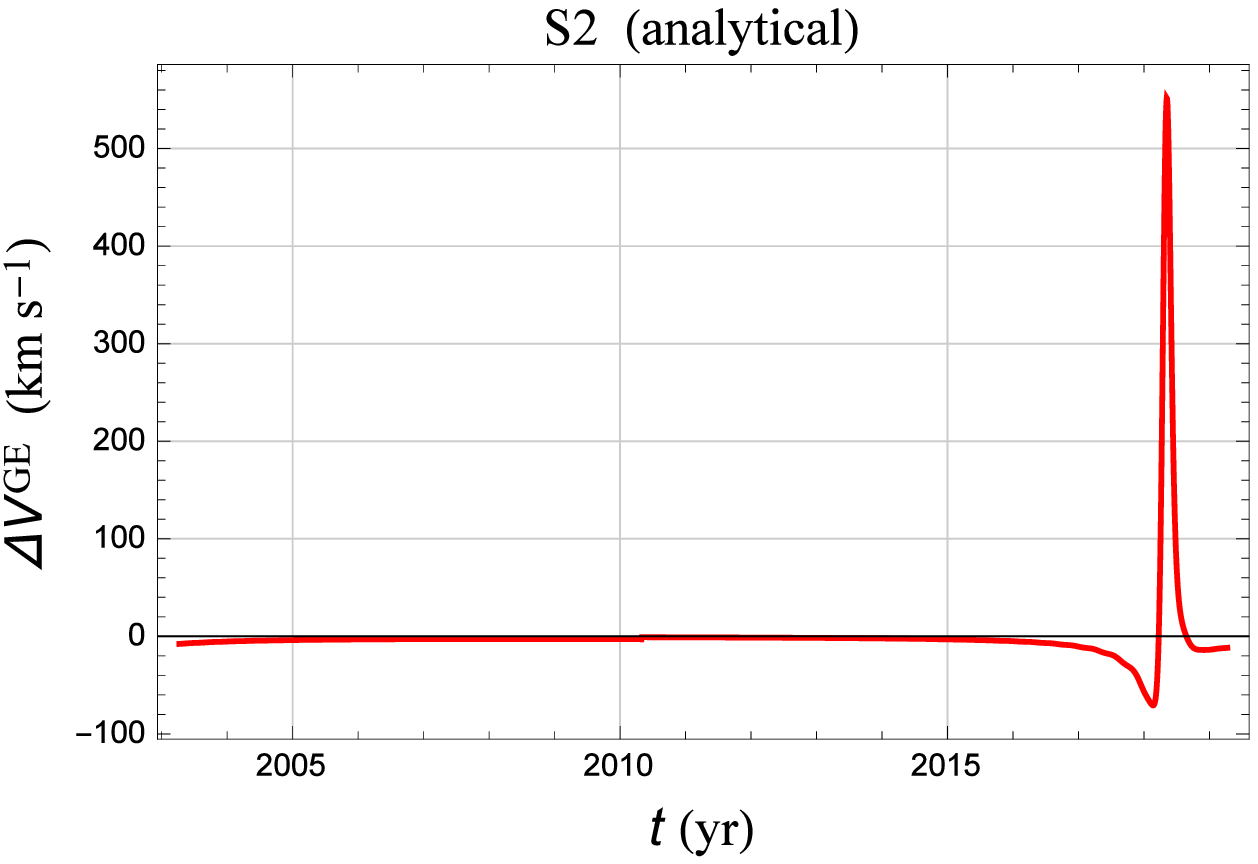}& \epsfysize= 5.00 cm\epsfbox{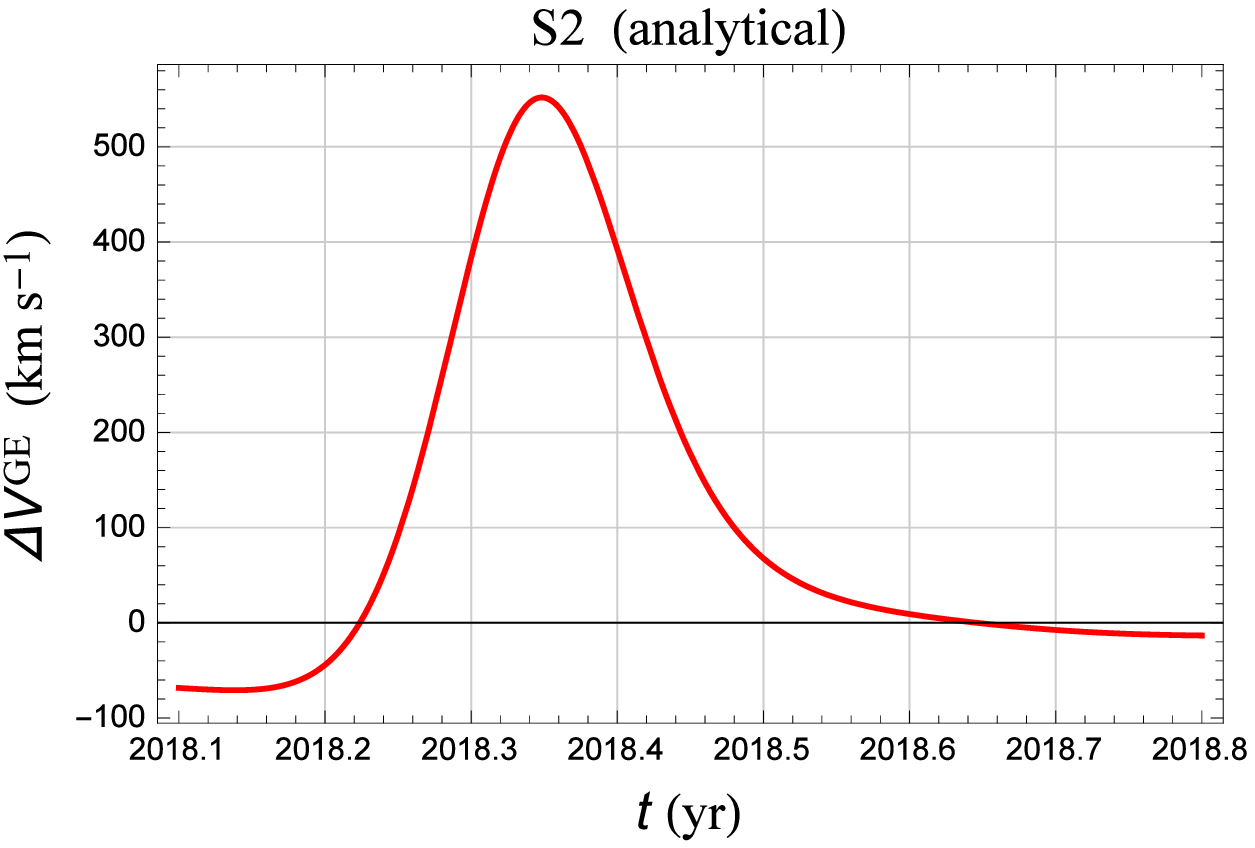}\\
\end{tabular}
}
}
\caption{Upper row, blue curve: \textcolor{black}{nominal} $\Delta V^\textrm{GE}\ton{t}$, in $\textrm{km~s}^{-1}$, of S2 as the outcome of the difference between two numerical integrations of the equations of motion in Cartesian coordinates over a time span ranging from $t_0=2003.271$, which corresponds to the beginning of the radial velocity measurements used in table 5 of \citet{2017ApJ...837...30G}, to $t_0+\Pb$. Both the integrations share the same (Keplerian) initial conditions for $f_0 = 139.72~\textrm{deg}$, corresponding to $t_0=2003.271$, \textcolor{black}{retrieved from the central values of Table \ref{S2},} and differ by the 1pN Schwarzschild-like acceleration, which was purposely switched off in one of the two runs. Lower row, red curve: \textcolor{black}{nominal} $\Delta V^\textrm{GE}\ton{t}$, in $\textrm{km~s}^{-1}$, of S2 obtained from \rfrs{Dvzf}{fMt} and the instantaneous changes of the Keplerian orbital elements, not displayed in the text, induced by \rfr{AGE}. The maximum of the 1pN gravitoelectric radial velocity shift amounts to $\Delta V^\textrm{GE}_\textrm{max} = 551~\textrm{km~s}^{-1}$, and occurs at $t_\textrm{max} = 2018.35$. It turns out that the net shift after a full revolution starting at $t_0=2003.271$ amounts to $\ang{\Delta V^\textrm{GE}} = -11.6~\textrm{km~s}^{-1}$. \textcolor{black}{It turns out that the numerical (blue) and the analytical (red) time series displayed here agree well within the RV measurement errors, with a maximum discrepancy of $\left|\delta\Delta V^\textrm{GE}\right|\lesssim \textcolor{black}{2-3}~\textrm{km~s}^{-1}$ in the range $2018.30-2018.40$ vanishing at $2018.35$: cfr. with Figure \ref{referee}.}
}\label{figura1}
\end{figure*}

\begin{figure*}
\centerline{
\vbox{
\begin{tabular}{cc}
\epsfysize= 5.00 cm\epsfbox{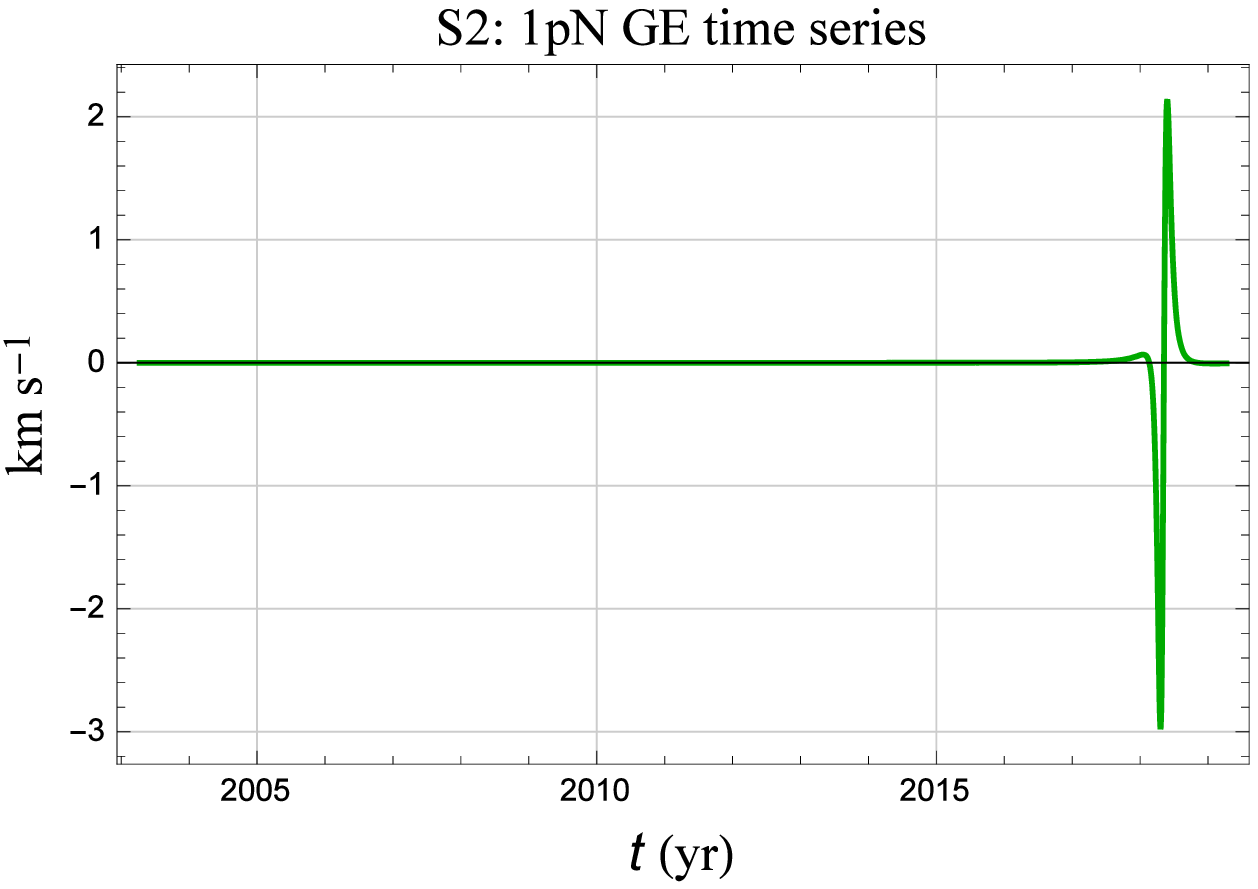}&\epsfysize= 5.00 cm\epsfbox{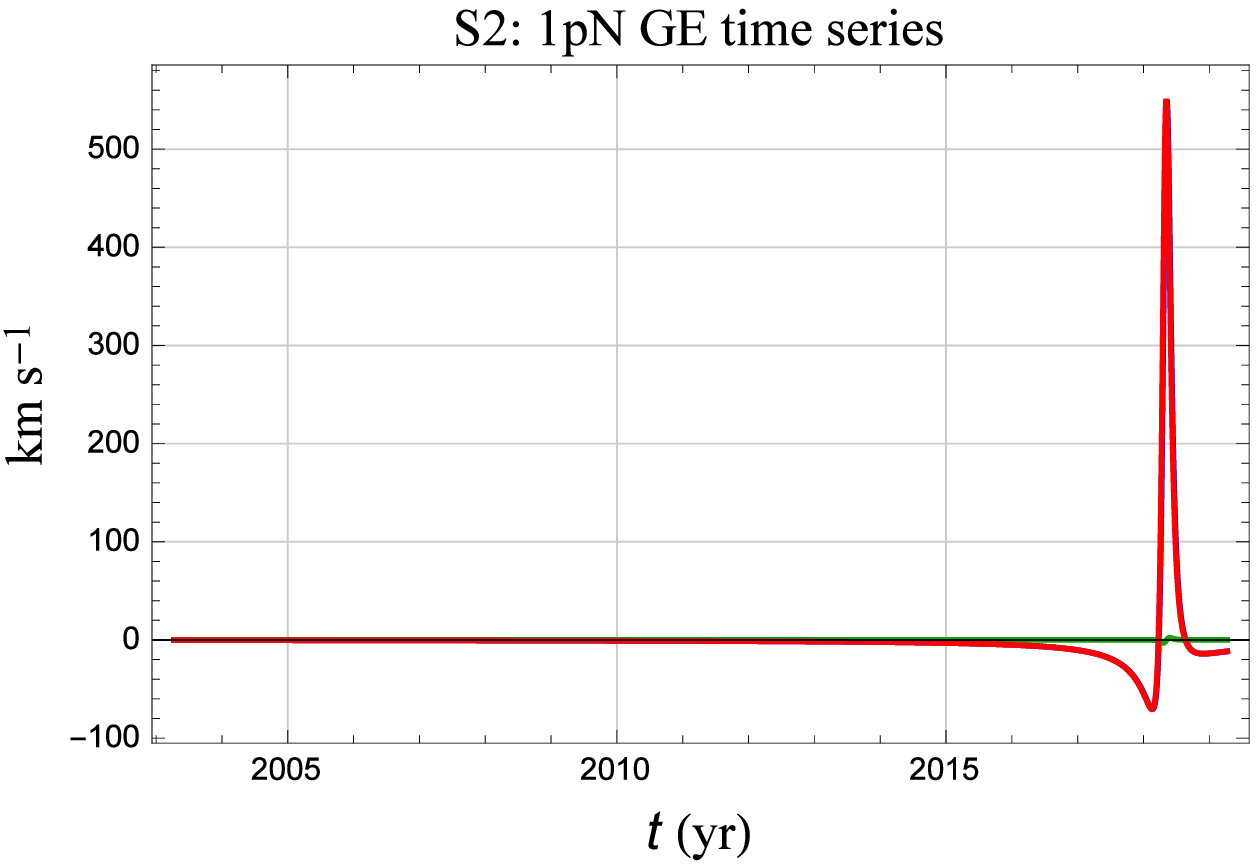}\\
\end{tabular}
}
}
\caption{\textcolor{black}{Green curve: difference $\delta\Delta V^\textrm{GE}\ton{t}$ between the numerical (blue) and the analytical (red) 1pN GE time series of the RV of S2 displayed separately in Figure \ref{figura1} \textcolor{black}{and superimposed in the right panel of the present Figure}. It turns out that $\delta\Delta V^\textrm{GE}\ton{t}$ vanishes for $t_\textrm{max}=2018.35$, reaching a maximum of about $\left|\delta\Delta V^\textrm{GE}\right|\lesssim \textcolor{black}{2-3}~\textrm{km~s}^{-1}$ at $2018.30$ and $2018.40$. Outside such a range, it goes rapidly to zero. The discrepancy between the numerical and the analytical time series around the periastron passage is not statistically significative, given the size of the RV measurement errors.}
}\label{referee}
\end{figure*}

\begin{figure*}
\centerline{
\vbox{
\begin{tabular}{cc}
\epsfysize= 5.00 cm\epsfbox{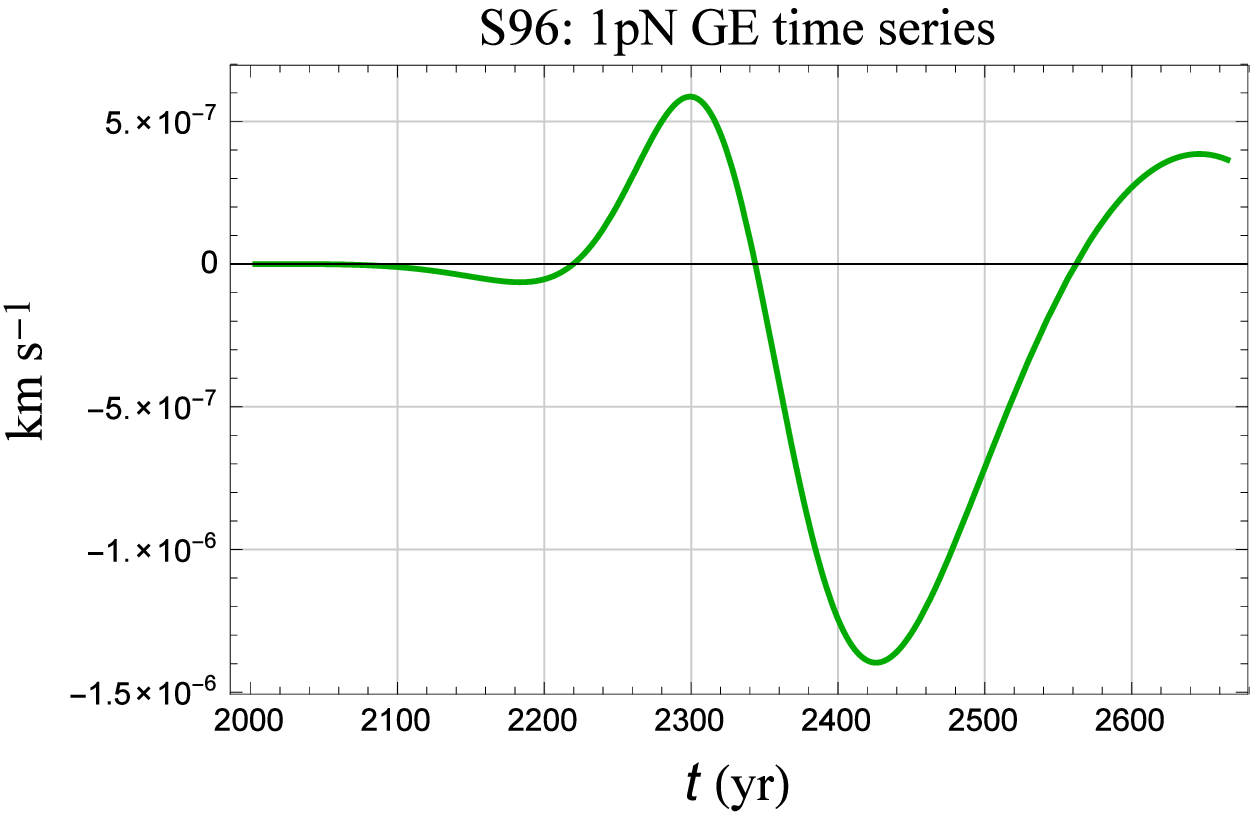}&\epsfysize= 5.00 cm\epsfbox{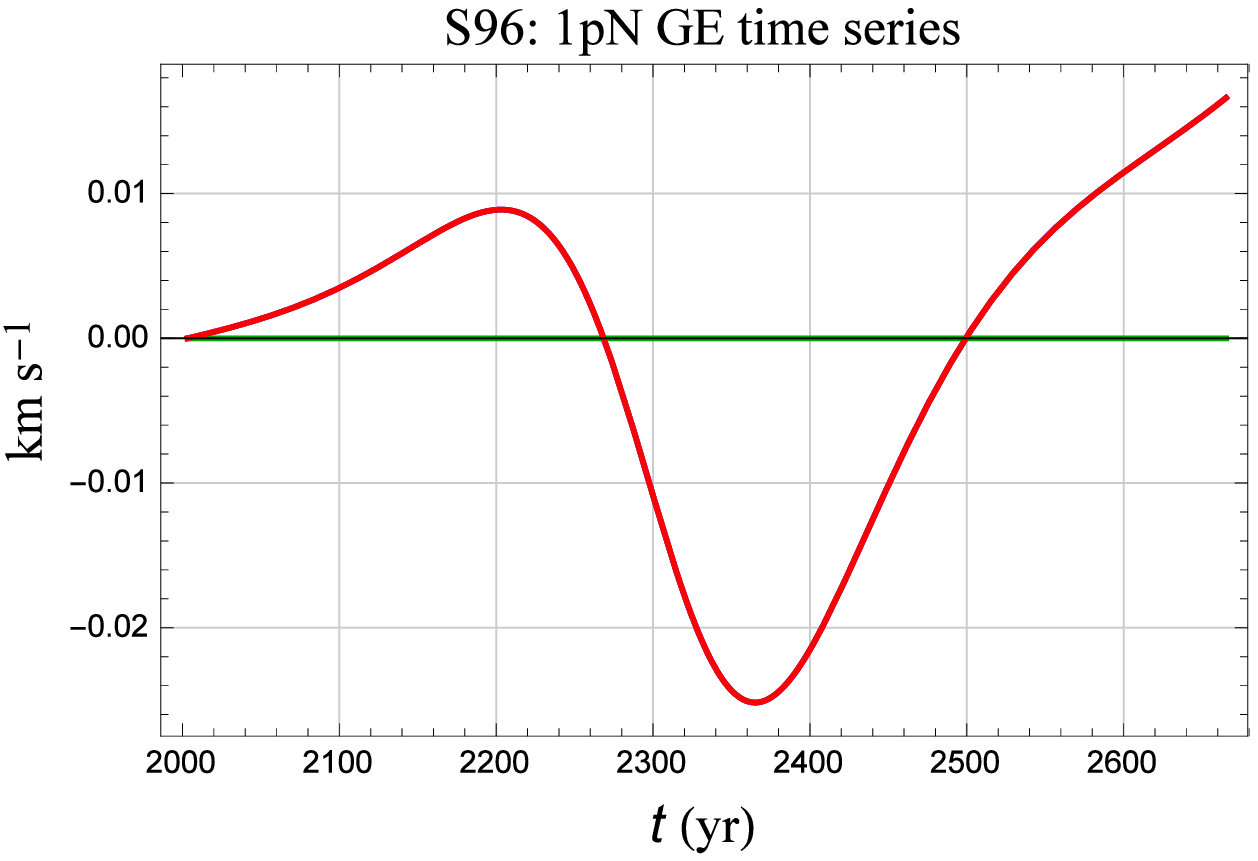}\\
\epsfysize= 5.00 cm\epsfbox{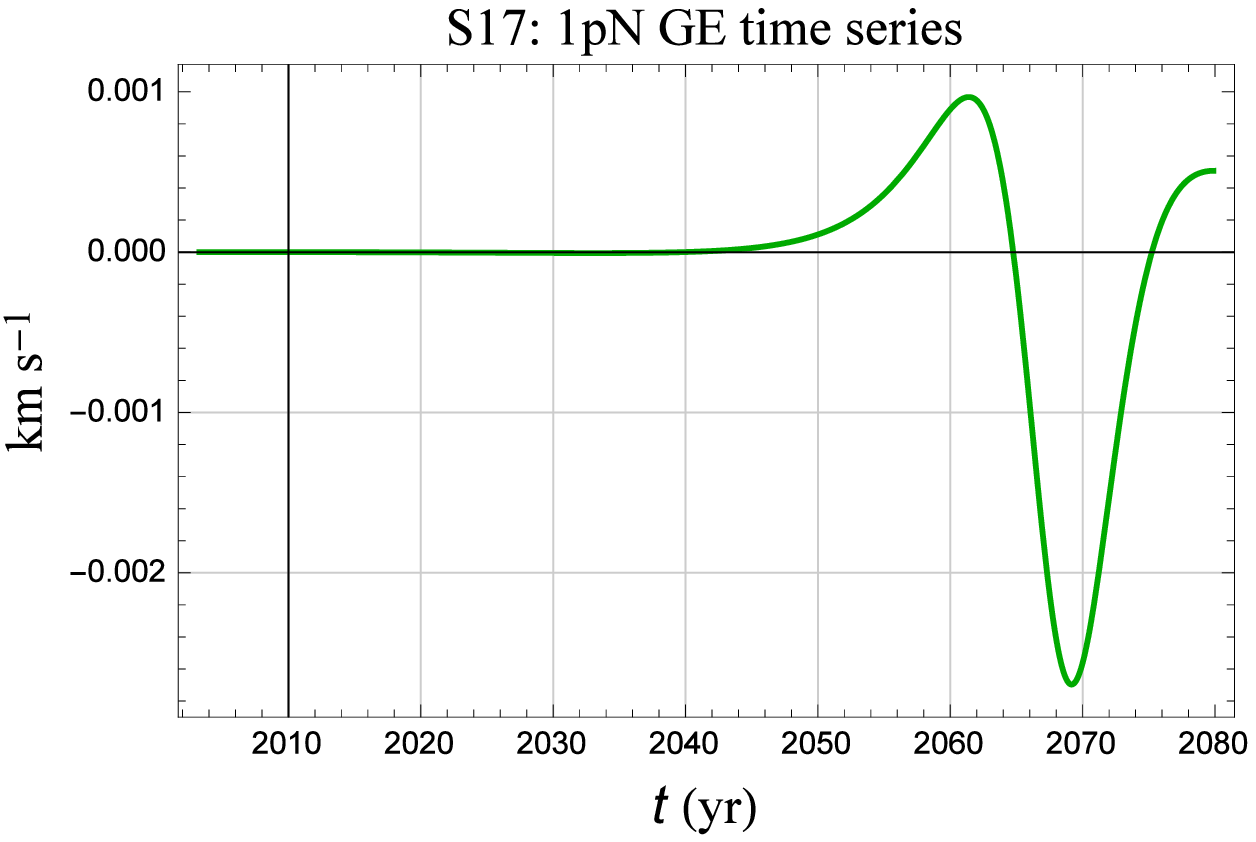}&\epsfysize= 5.00 cm\epsfbox{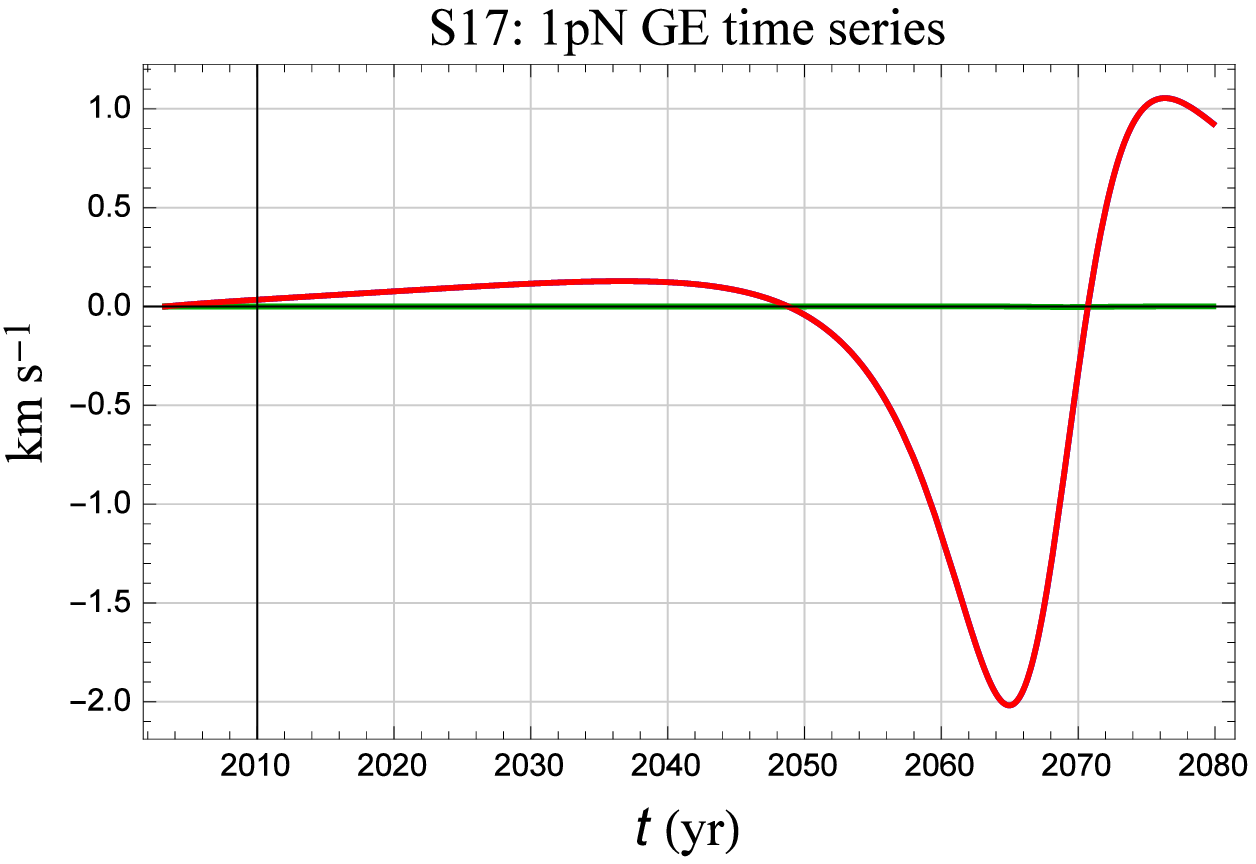}\\
\epsfysize= 5.00 cm\epsfbox{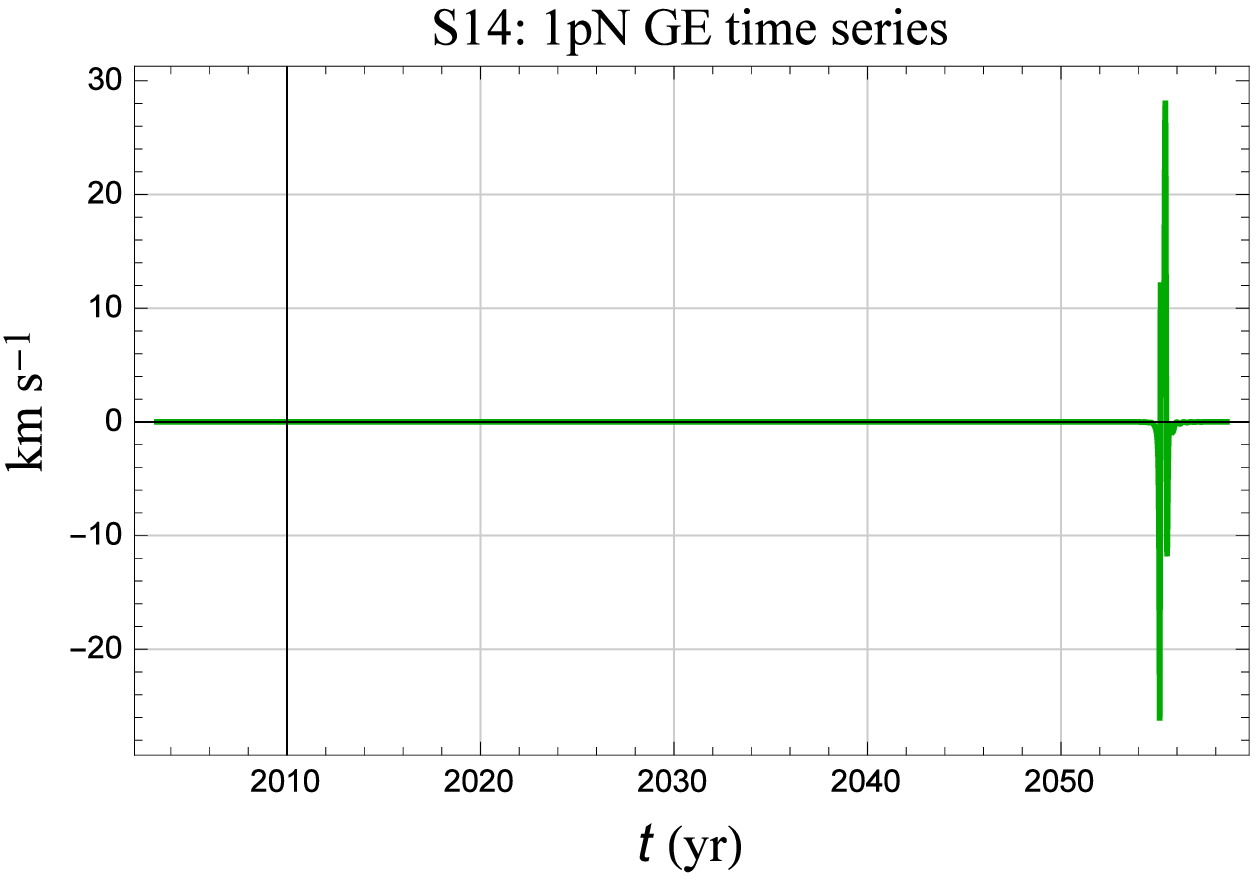}&\epsfysize= 5.00 cm\epsfbox{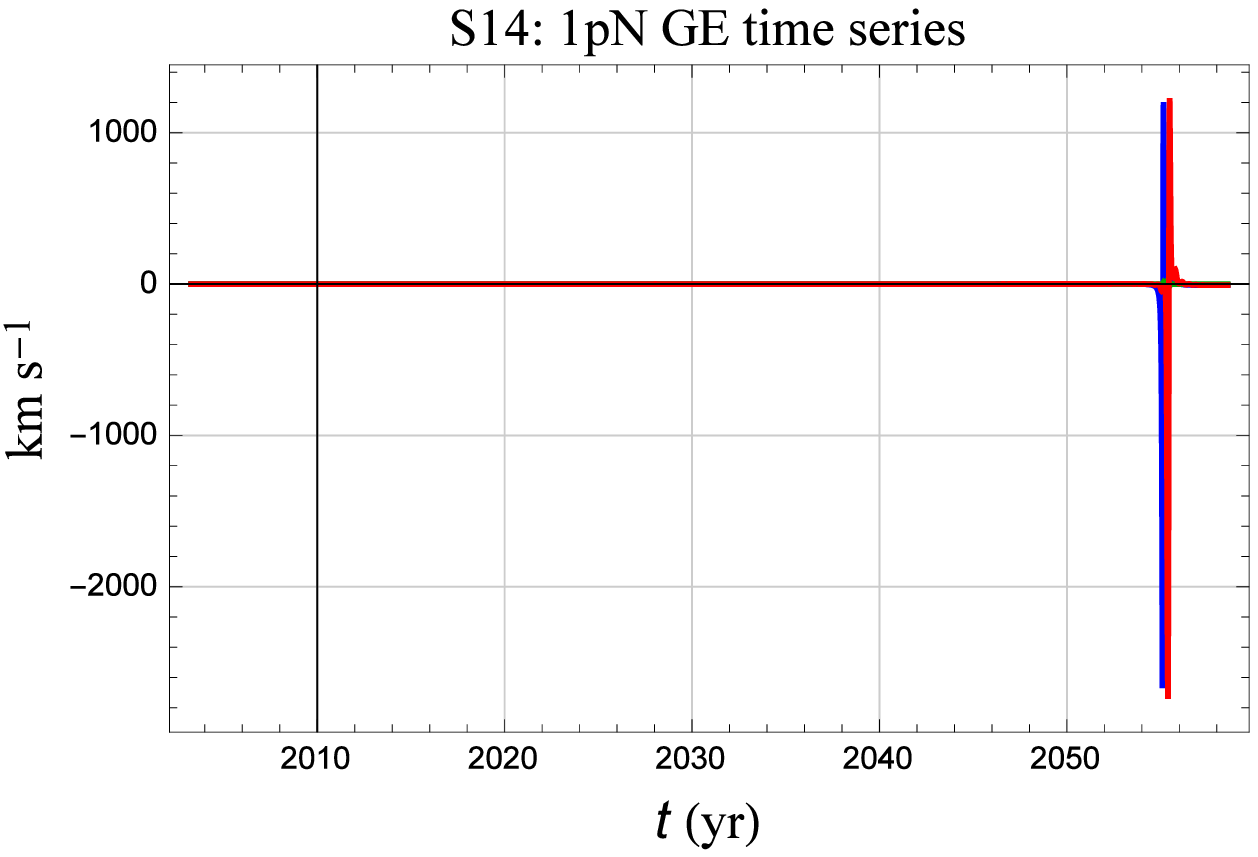}\\
\end{tabular}
}
}
\caption{\textcolor{black}{Differences (green curves) between the numerically integrated (blue curves) and the analytically calculated (red curves) time series of the 1pN GE shifts of the RV of S96 ($e = 0.174$), S17 ($e = 0.397$), S14 ($e = 0.9761$) obtained for $s_\textrm{max}=j_\textrm{max}=350$ in \rfr{fMt}.
}  }\label{referee2}
\end{figure*}

For $t_0=2003.271$, which is the starting epoch of the RV measurements collected in table 5 of \citet{2017ApJ...837...30G}, it is
\eqi\ang{\Delta V^\textrm{GE}} = -11.6~\textrm{km~s}^{-1}.\eqf
About the currently available RV measurements of S2, which  will cover a full orbital revolution in the next couple of years by including also the passage at periastron in 2018.33, Figure \ref{figura1} displays the corresponding \textcolor{black}{nominal} temporal pattern of the 1pN gravitoelectric RV signal according to either our analytical results and a numerical integration of the equations of motion, \textcolor{black}{both starting from the central values of the stellar and SMBH's  orbital and physical parameters listed in Table \ref{S2}}; its main quantitative features are summarized in Table \ref{resumeRV}. \textcolor{black}{Figure \ref{referee} displays the difference between our numerical and analytical time series: they agree well within the measurement errors in the S2's RV. In particular, the height of the peak is the same in both cases, being the corresponding epoch shifted by an amount within the uncertainty level of the time-type parameters of S2. The discrepancy of up to $\sim \textcolor{black}{2-3}~\textrm{km~s}^{-1}$ occurring in a narrow temporal range around the peak's epoch is not statistically significative. Figure \ref{referee2} shows that \textcolor{black}{the good agreement between our analytical calculation and numerical integrations hold also for higher values of the orbital eccentricity}.
} From Figure \ref{figura1} it turns out that the \textcolor{black}{nominal} 1pN RV signature, after staying quiet around the level of $\lesssim 10~\textrm{km~s}^{-1}$ for most of time, dramatically changes around the periastron passage by suddenly varying its amplitude from $-70~\textrm{km~s}^{-1}$ in 2018.13 to  $+551~\textrm{km~s}^{-1}$ in $t_\textrm{max} = 2018.35$. It may be interesting to note that the last available RV measurement in table 5 of \citet{2017ApJ...837...30G} was collected in  $t_\textrm{fin} = 2016.519$; it can be shown that the 1pN RV signature does not exceed  $-7~\textrm{km~s}^{-1}$ from $t_0$ to $t_\textrm{fin}$. Incidentally, this explains why general relativity was not yet measured so far; indeed, the published uncertainties $\sigma_{V_\textrm{S2}}$ in the RV measurements of S2 are of the order of  $\simeq 30-50~\textrm{km~s}^{-1}$ \citet{2017ApJ...837...30G}. \textcolor{black}{As far as the impact of the errors in the parameters of Table \ref{S2} on the 1pN RV signal is concerned, they change both its peak amplitude $\Delta V^\textrm{GE}_\textrm{max}$ and the corresponding epoch $t_\textrm{max}$ to a certain extent. It turns out that the most effective uncertainties are those on $\Pb,~e,~t_p,~\mu$ which shift the 1pN peak from a minimum of $\sim 505~\textrm{km~s}^{-1}$ in $2018.79$ to a maximum of $\sim 575 = ~\textrm{km~s}^{-1}$ in $2018.45$. Instead, the uncertainties in $I,~\Omega,~\omega$ do not play an appreciable role.} Our \textcolor{black}{preliminary} analysis shows that an accurate monitoring of the RV curve of S2 around the time of periastron passage occurring next year may allow for a detection of the general relativistic Schwarzschild-like  component of the SMBH's field.
\textcolor{black}{In fact, also the effects on the propagation of the electromagnetic waves (e.g. R{\o}mer time delay, transverse Doppler shift, gravitational redshift and lensing) to various pN orders and within full general relativity should be taken into account in a comprehensive analysis devoted to the actual measurability of the pN field of the SMBH with S2 next year. They should deserve further, dedicated analyses which are outside the scopes of the present one. Earlier studies, not focussed on the epoch considered here, can be found, e.g., in \citet{2006ApJ...639L..21Z,2010ApJ...711..157A}. }

About the traditional view of testing pN gravity with the periastron precession \textcolor{black}{(see e.g. \citet{2001A&A...374...95R, 2005ApJ...622..878W,2009ApJ...703.1743P})}, it could not work for S2, at least in the next few years. Indeed, from \rfr{Domega} it turns out that its net advance per orbit, which is independent of $f_0$, is as little as
\eqi
\ang{\Delta\omega^\textrm{GE}}=0.2~\textrm{deg}\lb{piccio}
\eqf
for S2.
On the other hand, according to table 3 of \citet{2017ApJ...837...30G}, the current accuracy in estimating the periastron of S2 amounts to just
\eqi
\sigma_{\omega_\textrm{S2}} = 0.57~\textrm{deg}.
\eqf
Thus, it is difficult to think that the 1pN relativistic periastron advance of S2 could be detectable before the completion of at least 3 or 4 full orbital revolutions. \textcolor{black}{Nonetheless, the situation may become more favorable with the new interferometric facility GRAVITY\footnote{\textcolor{black}{See http://www.mpe.mpg.de/ir/gravity on the Internet.}} \citep{GRAVITYAA2017}.}
\section{The 1pN gravitomagnetic Lense-Thirring effect}\lb{LTeffects}
The stationary component of the 1pN field, due to mass-energy currents, is responsible of several aspects of the so-called spin-orbit coupling, or frame-dragging \citep{1986SvPhU..29..215D,1988nznf.conf..573T,2004GReGr..36.2223S,2009SSRv..148...37S}.

The 1pN gravitomagnetic, Lense-Thirring-type, acceleration affecting the relative orbital motion of a binary is \citep{1975PhRvD..12..329B,1989racm.book.....S}
\eqi
{\bds A}_\textrm{LT} = \rp{2G}{c^2 r^3}\qua{ 3\ton{\bds{\mathcal{S}}\bds\cdot\uR}\uR\bds\times\mathbf{v}  + \mathbf{v}\bds\times\bds{\mathcal{S}}  }.\lb{ALT}
\eqf
In general, it is
\eqi \kA\neq\kB,\eqf i.e. the angular momenta of the two bodies are usually not aligned. Furthermore, they are neither aligned with the orbital angular momentum $\bds L$, whose unit vector is given by $\uN$. Finally, also the magnitudes $S^\textrm{A},~S^\textrm{B}$ are, in general, different.

The radial, transverse and out-of-plane components of the gravitomagnetic acceleration, obtained by projecting \rfr{ALT} onto the unit vectors
$\uR,~\uT,~\uN$, turn out to be
\begin{align}
A_{\rho}^\textrm{LT} \lb{ARLT} & = \rp{2G\nk \ton{1+e\csf}^4\bds{\mathcal{S}}\bds\cdot\uN}{c^2 a^2\ton{1-e^2}^{7/2}}, \\ \nonumber \\
A_{\tau}^\textrm{LT} \lb{ATLT} & = -\rp{2eG\nk \ton{1+e\csf}^3\snf~\bds{\mathcal{S}}\bds\cdot\uN}{c^2 a^2\ton{1-e^2}^{7/2}}, \\ \nonumber \\
A_{\nu}^\textrm{LT} \nonumber \lb{ANLT} & = -\rp{2G\nk\ton{1+e\csf}^3}{c^2 a^2 \ton{1-e^2}^{7/2}}\bds{\mathcal{S}}\bds\cdot\grf{\qua{e\co -\ton{2+3e\csf}\cu}\ul -\right.\\ \nonumber \\
&-\left. \rp{1}{2}\qua{e\so +4\su + 3e\sin\ton{\omega+2f}  }\um  }.
\end{align}

By using \rfrs{ARLT}{ANLT} in \rfr{Dk} and \rfrs{anom}{inte} and integrating them from $f_0$ to $f_0+2\uppi$, it is possible to straightforwardly calculate the 1pN gravitomagnetic net orbital changes for a generic binary arbitrarily oriented in space: they are
\begin{align}
\ang{\Delta a^\textrm{LT}} \lb{DaeMLT}& = \ang{\Delta e^\textrm{LT}}=\ang{\Delta\mathcal{M}^\textrm{LT}} = 0, \\ \nonumber \\
\ang{\Delta I^\textrm{LT}} \lb{DILT}& = \rp{4\uppi G\bds{\mathcal{S}}\bds\cdot\ul}{c^2 \nk a^3\ton{1-e^2}^{3/2}}, \\ \nonumber \\
\ang{\Delta \Omega^\textrm{LT}} \lb{DOLT}& = \rp{4\uppi G\csc I\bds{\mathcal{S}}\bds\cdot\um}{c^2 \nk a^3\ton{1-e^2}^{3/2}}, \\ \nonumber \\
\ang{\Delta \omega^\textrm{LT}} \lb{DoLT}& = -\rp{4\uppi G\bds{\mathcal{S}}\bds\cdot\ton{2\uN + \cot I\um}}{c^2 \nk a^3\ton{1-e^2}^{3/2}}.
\end{align}
It is interesting to remark that, in the case of \rfr{ALT}, both \rfr{detadt} and \rfr{inte} yield vanishing contributions to $\ang{\Delta\mathcal{M}^\textrm{LT}}$.
For previous calculations based on different approaches and formalisms, see, e.g.,
\citet{1975PhRvD..12..329B, 1988NCimB.101..127D,2008ApJ...674L..25W,2009ApJ...703.1743P,2011PhRvD..84l4001I}, and references therein.

\Rfr{Dvzf}, calculated with \rfrs{DaeMLT}{DoLT}, allows to obtain the net change per revolution of the radial velocity of the visible binary's component A as
\begin{align}
\ang{\Delta V_\textrm{A}^\textrm{LT}} \lb{dvzLT} & = \rp{4\uppi G\mB}{c^2 M_\textrm{tot} p^2}\bds{\mathcal{S}}\bds\cdot\grf{ \ton{2\uN\sI+\um\cI}\ton{e\so+\sin u_0}  +\ul\cI\ton{ e\co+\cos u_0 }     }.
\end{align}
The analytical expression of the instantaneous shift $\Delta V_\textrm{A}^\textrm{LT}\ton{f}$ will not be displayed here since it is rather cumbersome.
From \rfr{dvzLT} it can be noted that, in general, the angular momenta of both the binary's components A,~B affect the RV of the spectroscopically detectable body A.

\begin{figure*}
\centerline{
\vbox{
\begin{tabular}{cc}
\epsfysize= 5.00 cm\epsfbox{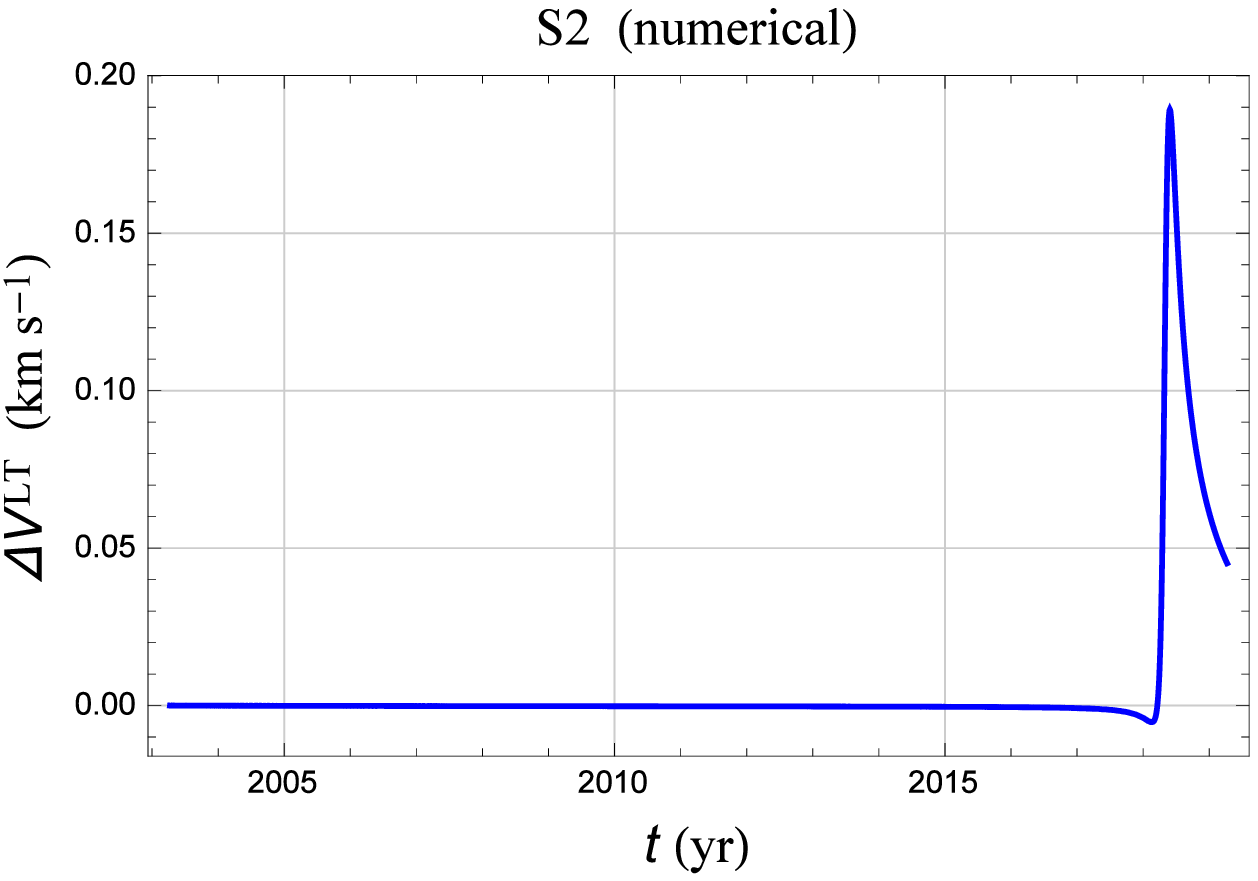}&\epsfysize= 5.00 cm\epsfbox{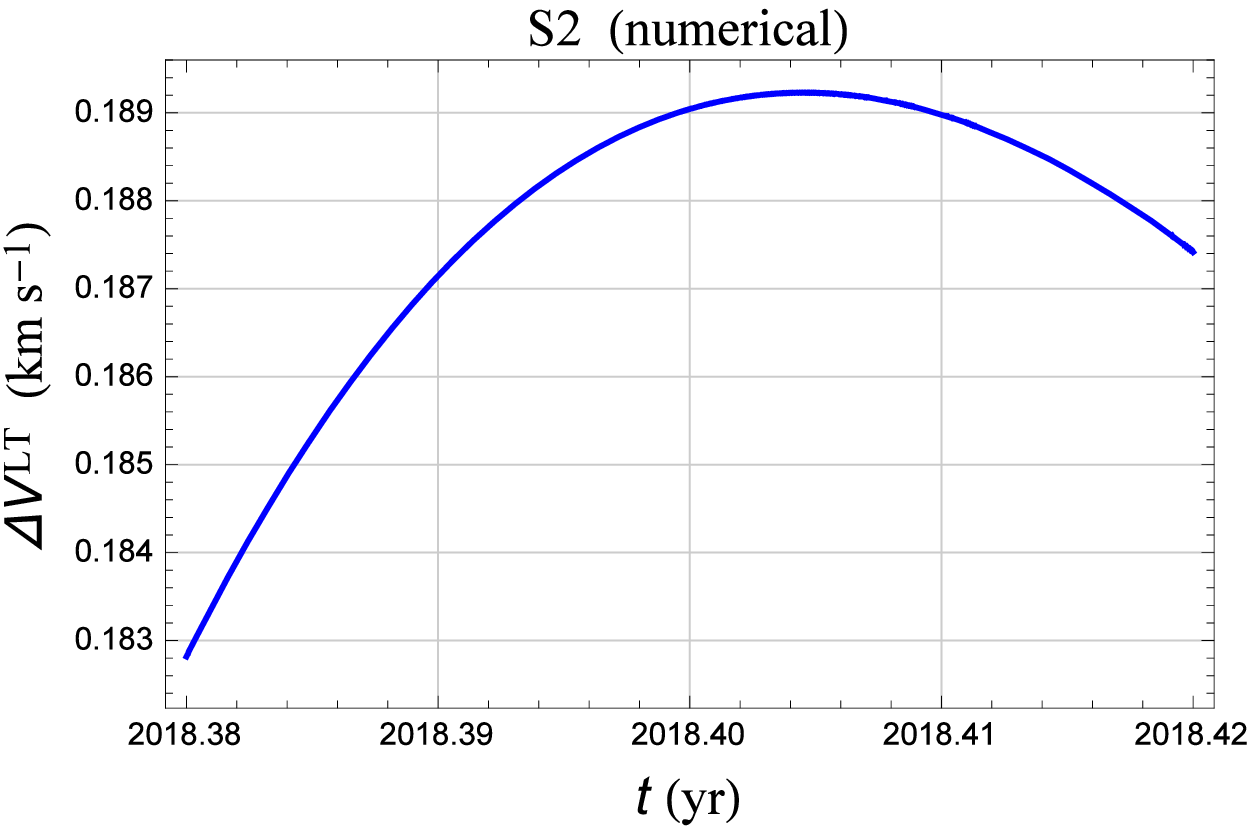}\\
\epsfysize= 5.00 cm\epsfbox{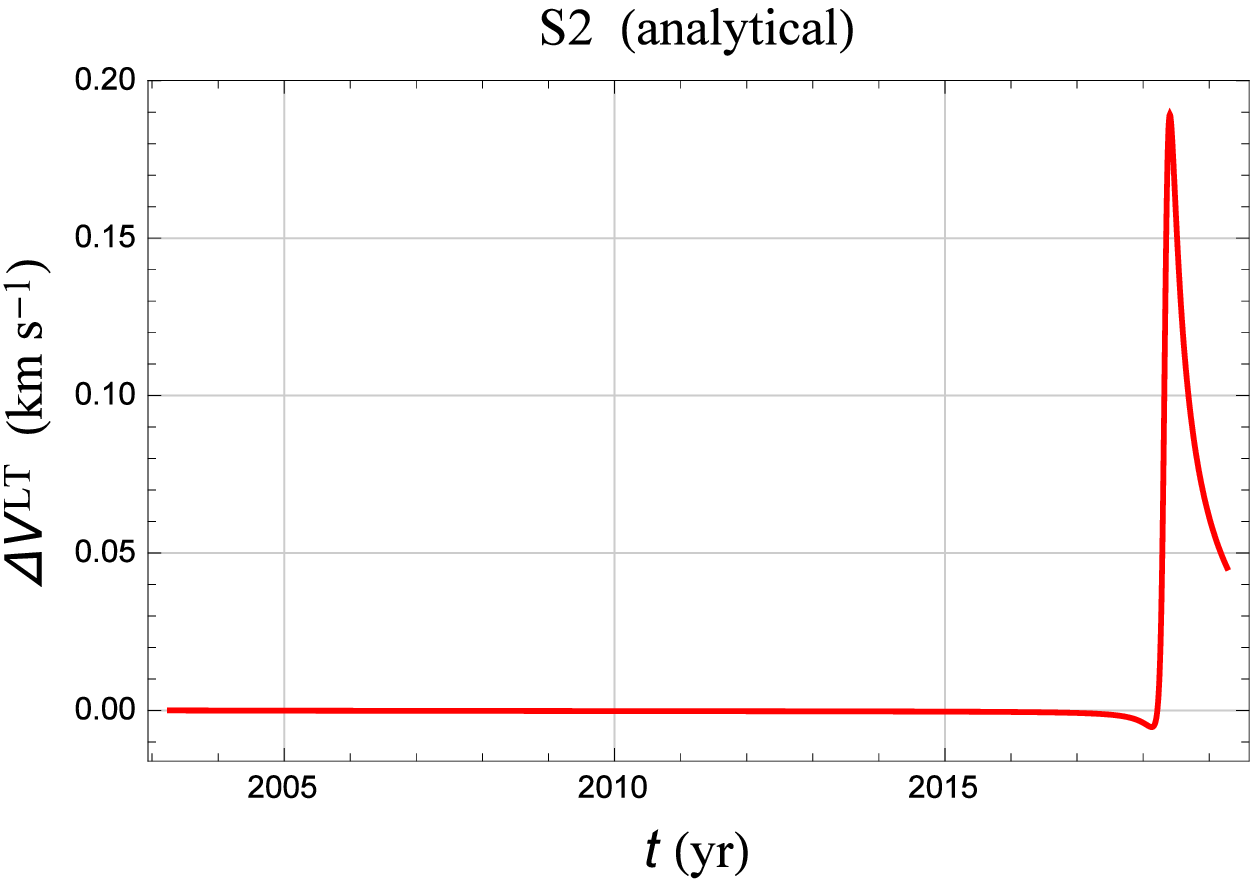}& \epsfysize= 5.00 cm\epsfbox{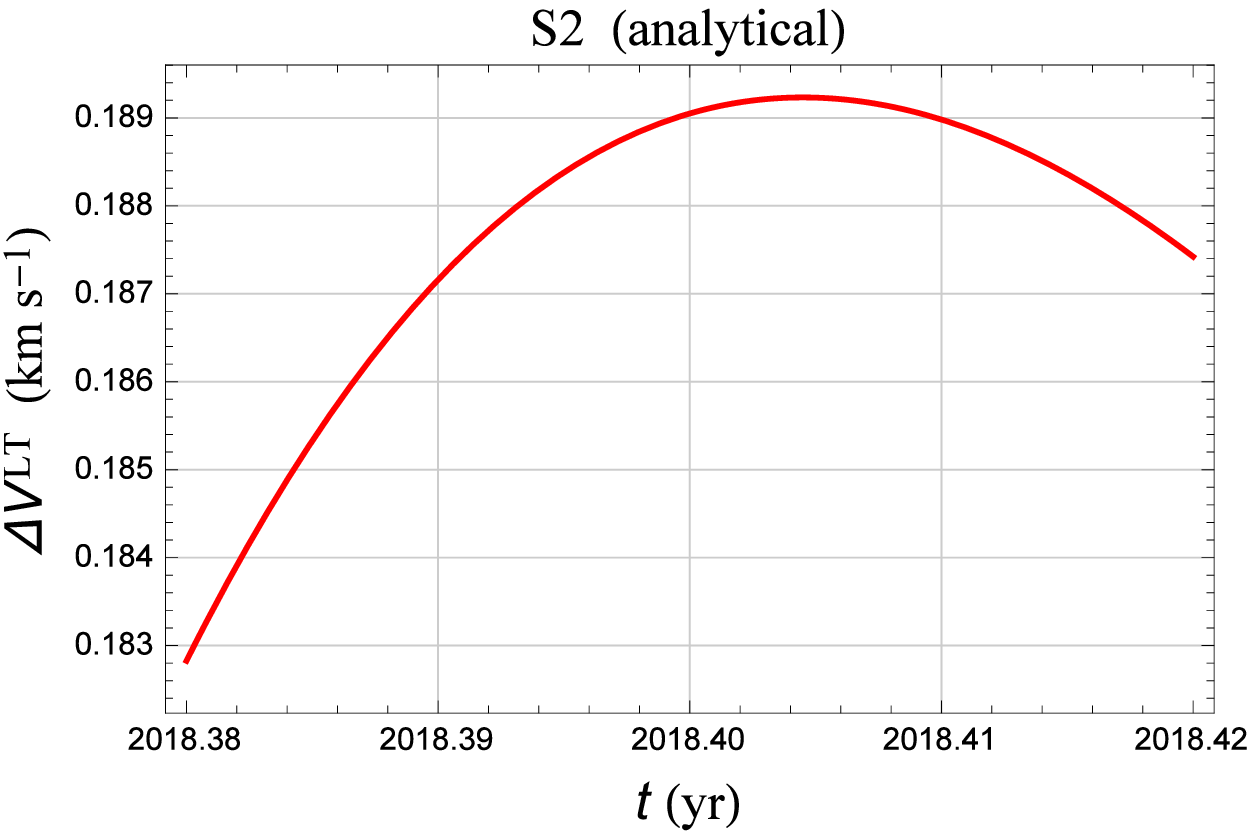}\\
\end{tabular}
}
}
\caption{Upper row, blue curve: \textcolor{black}{nominal} $\Delta V^\textrm{LT}\ton{t}$, in $\textrm{km~s}^{-1}$, of S2 as the outcome of the difference between two numerical integrations of the equations of motion in Cartesian coordinates over a time span ranging from $t_0=2003.271$, which corresponds to the beginning of the radial velocity measurements use in table 5 of \citet{2017ApJ...837...30G}, to $t_0+\Pb$. Both the integrations share the same (Keplerian) initial conditions for $f_0 = 139.72~\textrm{deg}$, corresponding to $t_0=2003.271$, \textcolor{black}{retrieved from the central values of Table \ref{S2},} and differ by the 1pN Lense-Thirring acceleration, which was purposely switched off in one of the two runs. Lower row, red curve: \textcolor{black}{nominal} $\Delta V^\textrm{LT}\ton{t}$, in $\textrm{km~s}^{-1}$, of S2 obtained from \rfrs{Dvzf}{fMt} and the instantaneous changes of the Keplerian orbital elements, not displayed in the text, induced by \rfr{ALT}. In both cases, the values $i^\bullet_\textrm{max}= 160~\textrm{deg},~\varepsilon^\bullet_\textrm{max}= 123 ~\textrm{deg}$ were adopted for the SMBH's spin axis orientation\textcolor{black}{: as per Table \ref{resumeRV}, they correspond to the maximum of $\Delta V^\textrm{LT}\ton{t}$}. The maximum of the 1pN gravitomagnetic radial velocity shift amounts to $\Delta V^\textrm{LT}_\textrm{max} = 0.19~\textrm{km~s}^{-1}$, and occurs at $t_\textrm{max} = 2018.40$; cfr. with Table \ref{resumeRV}.
}\label{figura2}
\end{figure*}

\textcolor{black}{
As far as Sgr A$^\ast$ is concerned, the orientation of the SMBH's spin is required in order to predict the frame-dragging of the S2's RV. We model it as
\eqi
{\bds{\hat{S}}}^\bullet =\sin i_\bullet\cos\varepsilon_\bullet~\uI + \sin i_\bullet\sin\varepsilon_\bullet~\uJ +\cos i_\bullet~\uK.
\eqf
The angles $i_\bullet,~\varepsilon_\bullet$ are still rather poorly constrained \citep{2009ApJ...697...45B,2011ApJ...735..110B,2016ApJ...827..114Y}; thus, we prefer to treat them as free parameters by considering their full ranges of variation
\begin{align}
0\lb{ispin}&\leq i_\bullet \leq 180~\textrm{deg}, \\ \nonumber \\
0\lb{espin}&\leq \varepsilon_\bullet \leq 360~\textrm{deg}.
\end{align}
Unfortunately, the Lense-Thirring effect turns out to be}  negligible for S2, as shown by Table \ref{resumeRV} and Figure \ref{figura2}\textcolor{black}{, produced with the values of Table \ref{resumeRV} for $i_\bullet,~\varepsilon_\bullet$ corresponding to the maximum RV shift.} Indeed, the gravitomagnetic field contributes  at most $0.2~\textrm{km~s}^{-1}$ to the star's RV. \textcolor{black}{Analyses on some spin-induced effects on the propagation of the electromagnetic waves can be found, e.g., in \citet{2015ApJ...809..127Z,2016ApJ...827..114Y,2017ApJ...834..198Z}.}
\section{The quadrupole-induced effect}\lb{Qeffects}
If the two bodies of the binary system under consideration are axisymmetric about their spin axes $\bds{\hat{S}}^\textrm{A/B}$, a further non-central relative acceleration arises; it is  \citep{1975PhRvD..12..329B}
\eqi
\rp{2r^4}{3\mu }{\bds A}_{J_2} = J_2^\textrm{A}R_\textrm{A}^2\grf{\qua{5\ton{\Kr}^2 - 1}\uR - 2\ton{\Kr}\kA} +{\textrm{A}\leftrightarrows\textrm{B}},\lb{AJ2}
\eqf
in which the first even zonal parameter $J_2^\textrm{A/B}$ is dimensionless.
In the notation of \citet{1975PhRvD..12..329B}, their $J_2^\textrm{A/B}$ parameter is not dimensionless as ours, being dimensionally an area because it corresponds to our $J_2^\textrm{A/B} R_\textrm{A/B}^2$. Furthermore, \citet{1975PhRvD..12..329B} introduce an associated dimensional quadrupolar parameter $\Delta I^\textrm{A/B}$, having the dimensions of a moment of inertia, which is connected to our $J_2^\textrm{A/B}$ by
\eqi
J_2^\textrm{A/B} = \rp{\Delta I^\textrm{A/B}}{M_\textrm{A/B}~R^2_\textrm{A/B}}.
\eqf
Thus, $\Delta I^\textrm{A/B}$ corresponds to the dimensional quadrupolar parameter $Q_2^\textrm{A/B}$ customarily adopted when astrophysical compact objects like neutron stars and black holes are considered \citep{1999ApJ...512..282L,
2014PhRvD..89d4043W}, up to a minus sign, i.e.
\eqi
J_2^\textrm{A/B}=-\rp{Q^\textrm{A/B}_2}{M_\textrm{A/B}~R_\textrm{A/B}^2}.\lb{J2Q}
\eqf
Thus,
\rfr{AJ2} can be written as
\eqi
\rp{2r^4}{3G}{\bds A}_{Q_2} = {\mathcal{Q}}_2^\textrm{A}\grf{\qua{1 - 5\ton{\Kr}^2 }\uR + 2\ton{\Kr}\kA }+{\textrm{A}\leftrightarrows\textrm{B}}\lb{AQ2}.
\eqf

Projecting \rfr{AJ2} onto the radial, tranvserse  and out-of-plane  unit vectors  $\uR,~\uT,~\uN$ provides us with
\begin{align}
\rp{2a^4\ton{1-e^2}^4}{3\mu\ton{1+e\csf}^4}A_{\rho}^{J_2} \lb{ARJ2}& = J_2^\textrm{A}R_\textrm{A}^2\grf{3\qua{\cu\ton{\Kl} + \su\ton{\Km}}^2 - 1}+{\textrm{A}\leftrightarrows\textrm{B}}, \\ \nonumber \\
-\rp{a^4\ton{1-e^2}^4}{3\mu\ton{1+e\csf}^4}A_{\tau}^{J_2} \lb{ATJ2}\nonumber & = J_2^\textrm{A}R_\textrm{A}^2\qua{\cu\ton{\Kl} + \su\ton{\Km}}\qua{\cu\ton{\Km} - \su\ton{\Kl}}+\\ \nonumber \\
&+ {\textrm{A}\leftrightarrows\textrm{B}}, \\ \nonumber \\
-\rp{a^4\ton{1-e^2}^4}{3\mu\ton{1+e\csf}^4}A_{\nu}^{J_2} \lb{ANJ2}& = J_2^\textrm{A}R_\textrm{A}^2\qua{\cu\ton{\Kl} + \su\ton{\Km}}\ton{\Kn}+{\textrm{A}\leftrightarrows\textrm{B}}.
\end{align}

A straightforward consequence of \rfrs{ARJ2}{ANJ2} is the calculation of the net quadrupole-induced shifts per revolution of the Keplerian orbital elements by means of \rfr{Dk} and \rfrs{anom}{inte}, which turn out to be
\begin{align}
\ang{\Delta a^{J_2}}&=\ang{\Delta e^{J_2}}\lb{DaeJ2}=0, \\ \nonumber \\
-\rp{p^2}{3\uppi}\ang{\Delta I^{J_2}} \lb{DIJ2}& = J_2^\textrm{A}R^2_\textrm{A}\ton{\Kl}\ton{\Kn}+{\textrm{A}\leftrightarrows\textrm{B}}, \\ \nonumber \\
-\rp{p^2}{3\uppi}\ang{\Delta\Omega^{J_2}} \lb{DOJ2}& = J_2^\textrm{A}R^2_\textrm{A}\ton{\Km}\ton{\Kn}\csc I+{\textrm{A}\leftrightarrows\textrm{B}}, \\ \nonumber \\
\nonumber \rp{2p^2}{3\uppi}\ang{\Delta\omega^{J_2}} \lb{DoJ2}& = J_2^\textrm{A}R^2_\textrm{A}\grf{2 - 3\qua{\ton{\Kl}^2+\ton{\Km}^2}+2\ton{\Km}\ton{\Kn}\cot I}+ \\ \nonumber \\
&+ {\textrm{A}\leftrightarrows\textrm{B}}, \\ \nonumber \\
\rp{2a^2\ton{1-e^2}^3}{3\uppi\ton{1+e\cos f_0}^3}\ang{\Delta\mathcal{M}^{J_2}} \nonumber \lb{DMJ2}& = J_2^\textrm{A}R^2_\textrm{A}\grf{2 - 3\qua{\ton{\Kl}^2+\ton{\Km}^2} - \right.\\ \nonumber \\
\nonumber &-\left. 3\qua{ \ton{\Kl}^2-\ton{\Km}^2 }\cos 2u_0 - 6\ton{\Kl}\ton{\Km}\sin 2u_0 } +\\ \nonumber \\
& +{\textrm{A}\leftrightarrows\textrm{B}}.
\end{align}
\textcolor{black}{See also \citet{2008ApJ...674L..25W}.}
Also \rfr{DMJ2}, as \rfr{DMGE} for the Schwarzschild-like 1pN acceleration, is a novel result which amends the incorrect formulas  widely disseminated in the literature \citep{2004Tapleyetal,2005ormo.book.....R,2005som..book.....C,2008orbi.book.....X,2011PhRvD..84l4001I}; indeed, it turns out that, in the case of \rfr{AJ2},  \rfr{inte} does not vanish when integrated over a full orbital revolution. Furthermore, contrary to almost all of the other derivations existing in the literature, \rfr{DMJ2} is quite general since it holds for a two-body system with generic quadrupole mass moments arbitrarily oriented in space, and characterized by a general orbital configuration. The same remark holds also for \rfrs{DaeJ2}{DoJ2}; cfr. with the corresponding (correct) results by \citet{2011PhRvD..84l4001I} in the case of a test particle orbiting an oblate primary.

According to \rfr{Dvzf} and \rfrs{DaeJ2}{DMJ2}, the net change of the radial velocity of the body A after one orbital revolution is
\begin{align}
\rp{2M_\textrm{tot}a\ton{1-e^2}^5}{3\mB\uppi\nk}\ang{\Delta V_\textrm{A}^{J_2}} \lb{DvzJ2}\nonumber & = J_2^\textrm{A}R^2_\textrm{A}\ton{1+e\cos f_0}^5\sI\sin u_0\cdot\\ \nonumber \\
\nonumber &\cdot\grf{-2 + 3\qua{\ton{\Kl}^2 + \ton{\Km}^2} +3\qua{\ton{\Kl}^2 - \ton{\Km}^2}\cos 2u_0 + \right.\\ \nonumber \\
\nonumber & + \left.6\ton{\Kl}\ton{\Km}\sin 2u_0  } + \\ \nonumber \\
\nonumber & + J_2^\textrm{A}R^2_\textrm{A}\ton{1-e^2}^{5/2}\grf{ -2 + 3\qua{\ton{\Kl}^2 + \ton{\Km}^2} }\sI\ton{e\so+\sin u_0} -\\ \nonumber \\
\nonumber &-2J_2^\textrm{A}R^2_\textrm{A}\ton{1-e^2}^{5/2}\ton{\Kn}\cI\qua{ \ton{\Kl}\ton{e\co + \cos u_0} + \right.\\ \nonumber \\
& + \left. \ton{\Km}\ton{e\so + \sin u_0}  } +{\textrm{A}\leftrightarrows\textrm{B}}.
\end{align}
It can be shown that \rfr{DvzJ2} does not vanish to the order zero in $e$.
The analytical expression of $\Delta V^{J_2}_\textrm{A}\ton{f}$  is far too ponderous to be explicitly displayed here.
As for frame-dragging, \rfr{DvzJ2} shows that, also in this case, both the binary's components A,~B contribute to the RV of the body A for which the light curve is available.

\begin{figure*}
\centerline{
\vbox{
\begin{tabular}{cc}
\epsfysize= 5.00 cm\epsfbox{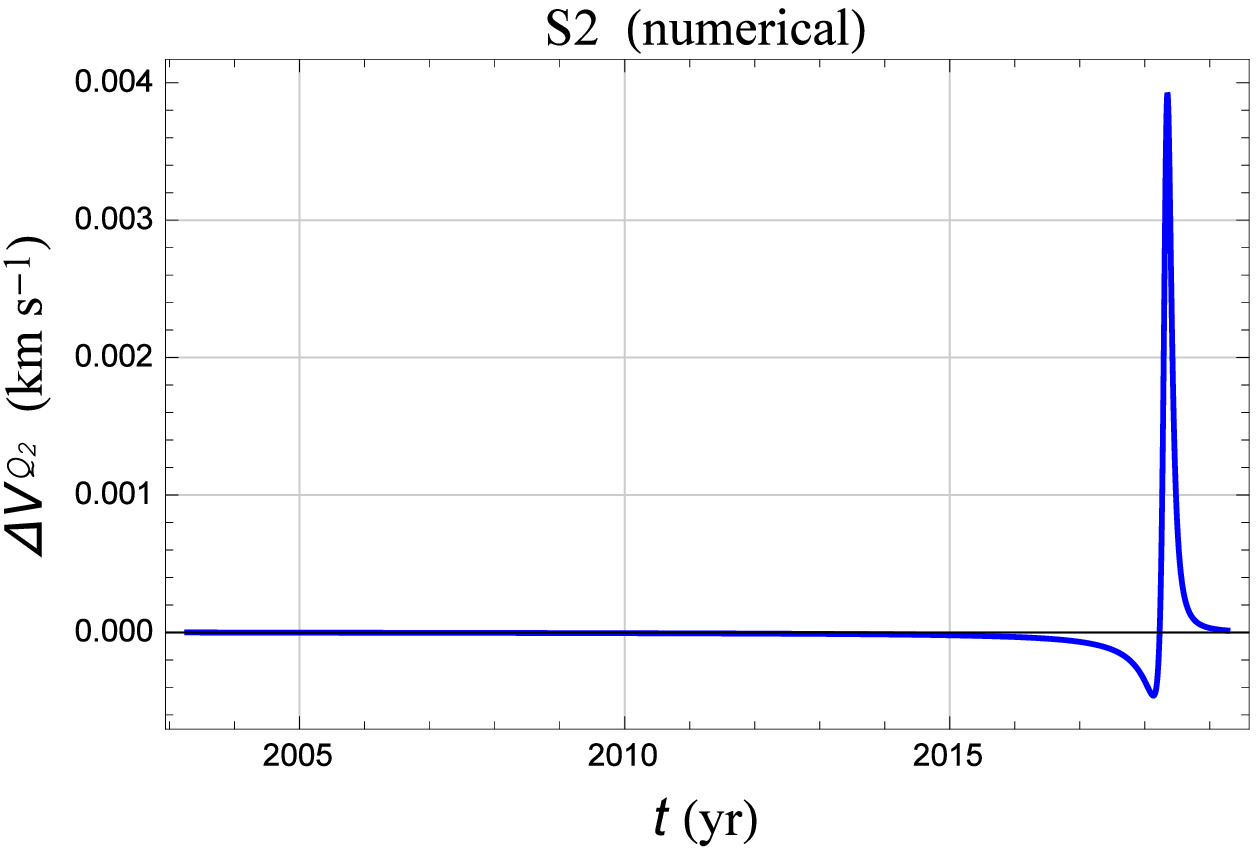}&\epsfysize= 5.00 cm\epsfbox{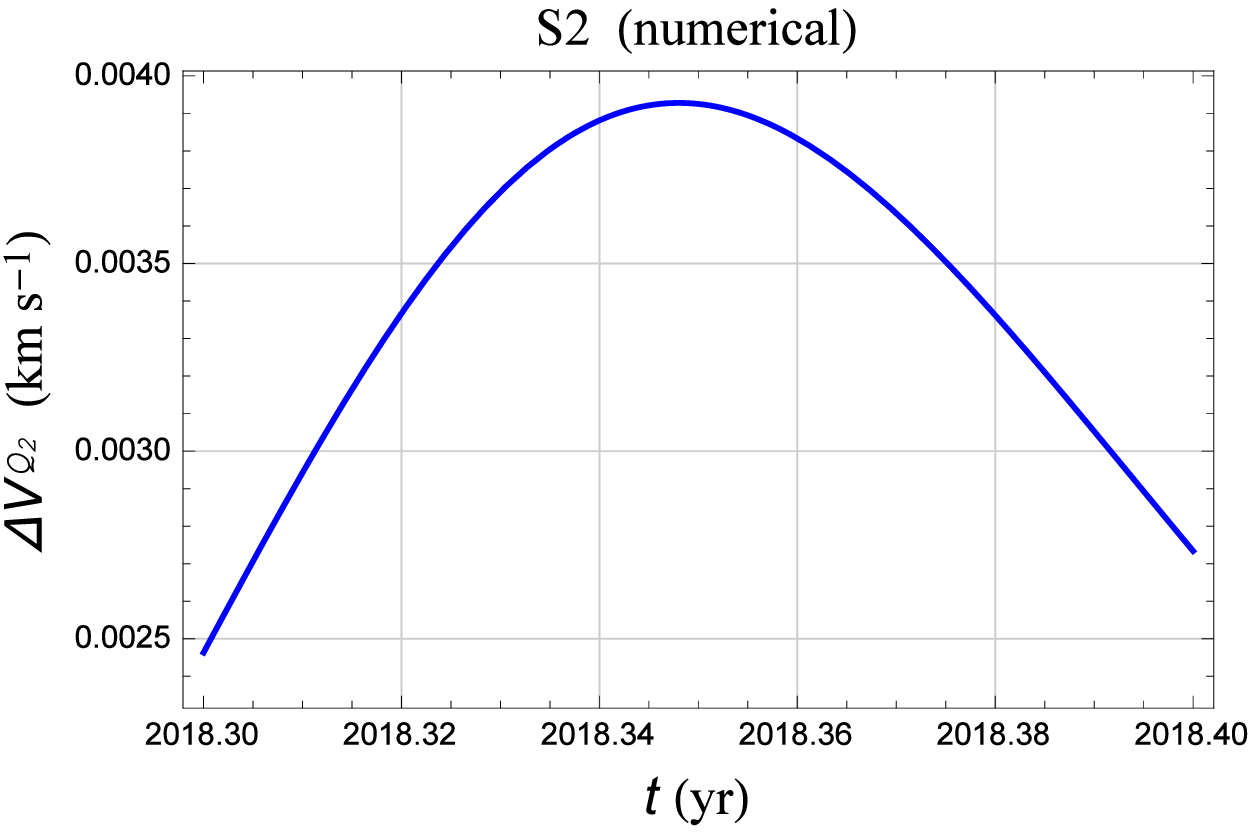}\\
\epsfysize= 5.00 cm\epsfbox{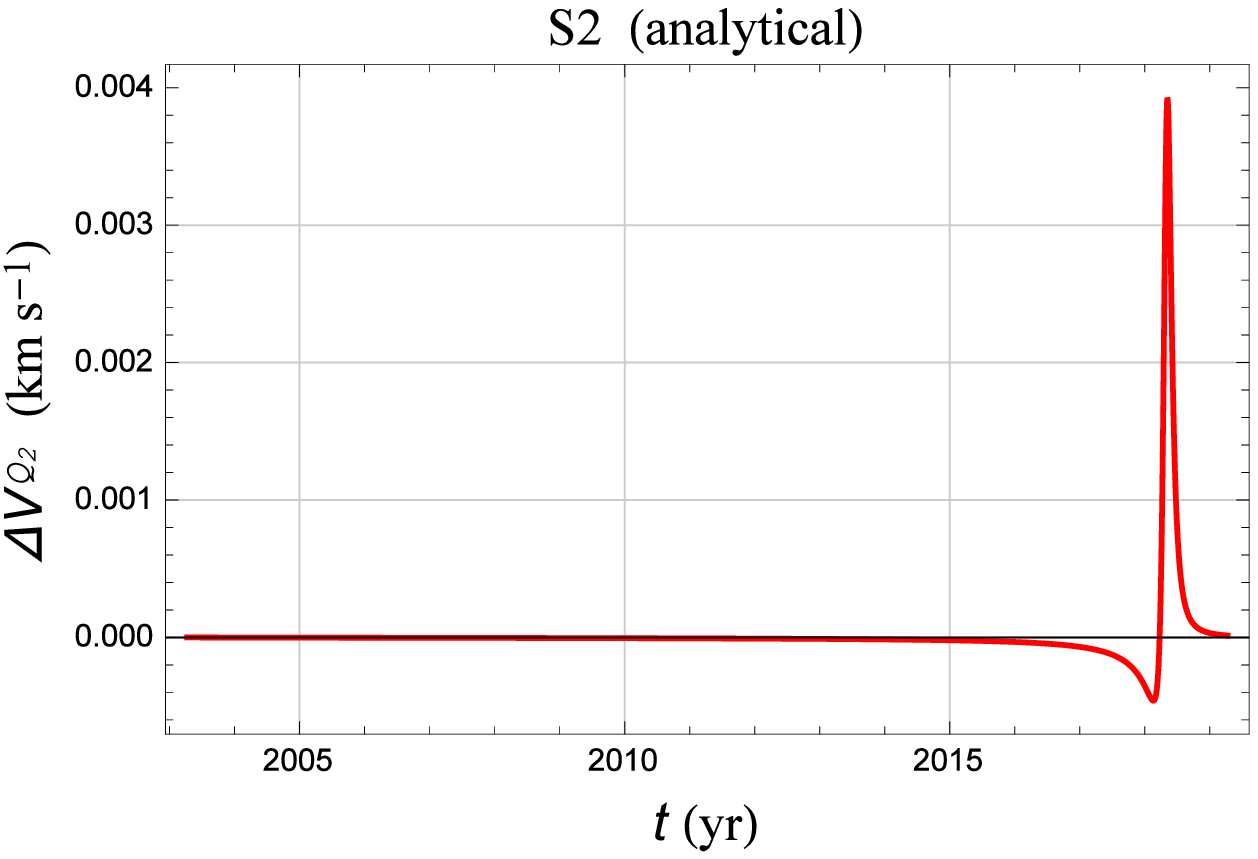}& \epsfysize= 5.00 cm\epsfbox{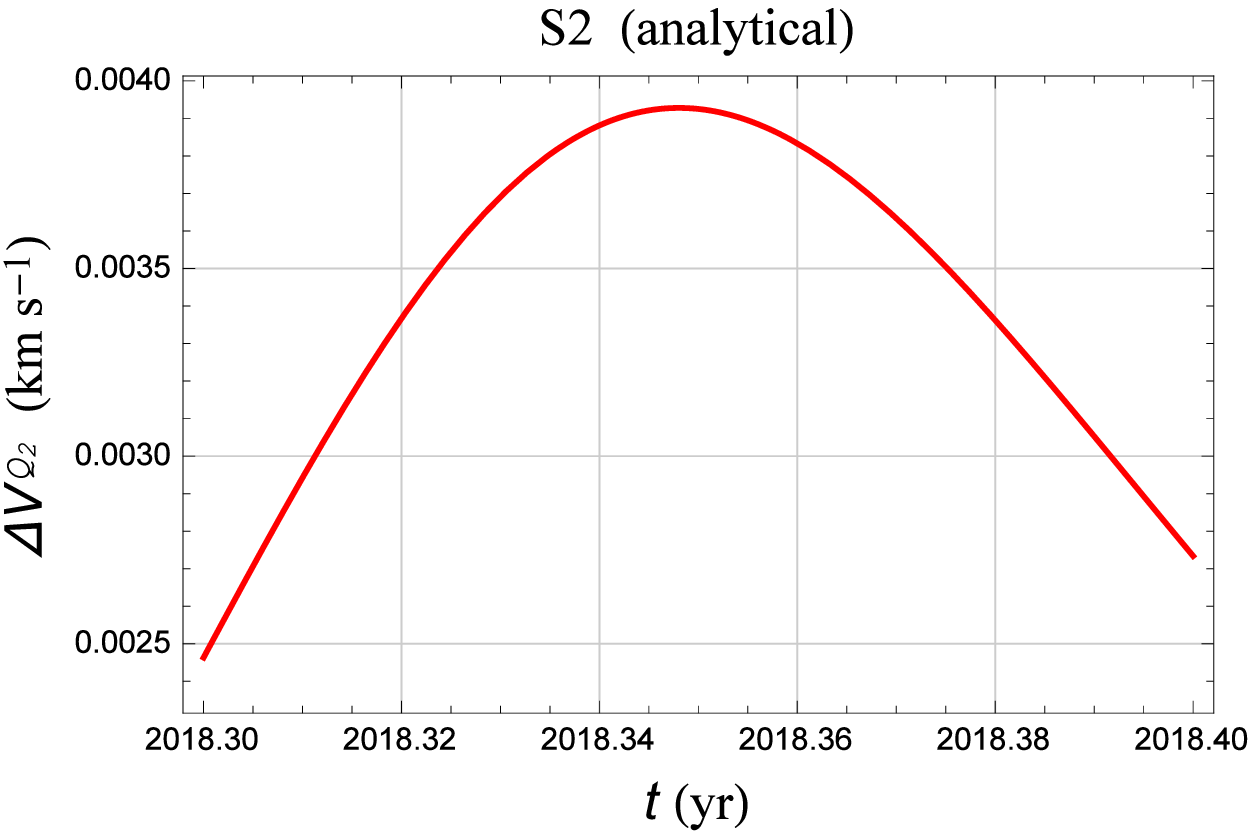}\\
\end{tabular}
}
}
\caption{Upper row, blue curve: \textcolor{black}{nominal} $\Delta V^{Q_2}\ton{t}$, in $\textrm{km~s}^{-1}$, of S2 as the outcome of the difference between two numerical integrations of the equations of motion in Cartesian coordinates over a time span ranging from $t_0=2003.271$, which corresponds to the beginning of the radial velocity measurements used in  table 5 of \citet{2017ApJ...837...30G}, to $t_0+\Pb$. Both the integrations share the same (Keplerian) initial conditions for $f_0 = 139.72~\textrm{deg}$, corresponding to $t_0=2003.271$, \textcolor{black}{retrieved from the central values of Table \ref{S2},} and differ by the quadrupole acceleration, which was purposely switched off in one of the two runs. Lower row, red curve: \textcolor{black}{nominal} $\Delta V^{Q_2}\ton{t}$, in $\textrm{km~s}^{-1}$, of S2 obtained from \rfrs{Dvzf}{fMt} and the instantaneous changes of the Keplerian orbital elements, not displayed in the text, induced by \rfr{AQ2}. In both cases, the values $i^\bullet_\textrm{max}= 72.9~\textrm{deg},~\varepsilon^\bullet_\textrm{max}= 207~\textrm{deg}$ were adopted for the SMBH's spin axis orientation\textcolor{black}{: as per Table \ref{resumeRV}, they correspond to the maximum of $\Delta V^{Q_2}\ton{t}$}. The maximum of the quadrupole-induced shift of the radial velocity amounts to $\Delta V^{Q_2}_\textrm{max} = 0.0039~\textrm{km~s}^{-1}$, and occurs at $t_\textrm{max} = 2018.348$; cfr. with Table \ref{resumeRV}.
}\label{figura3}
\end{figure*}

As far as S2 is concerned, the quadrupole of the SMBH has a completely negligible impact on its RV, as shown by Table \ref{resumeRV} and Figure \ref{figura3}\textcolor{black}{, obtained
 with the values of Table \ref{resumeRV} for $i_\bullet,~\varepsilon_\bullet$ yielding the maximum RV shift}. Indeed, its maximum value cannot be larger than $0.0039~\textrm{km~s}^{-1}$. \textcolor{black}{In, e.g., \citet{2015ApJ...809..127Z,2016ApJ...827..114Y,2017ApJ...834..198Z} some analyses on the SMBH's quadrupole impact on the propagation of electromagnetic waves can be found.}
\section{Summary and conclusions}\lb{fine}
The radial velocity $V$ is one of the key direct observables of a binary system whenever a spectroscopic light curve is available for at least one of its components. Thus, we devised a perturbative strategy to analytically calculate both the instantaneous change $\Delta V\ton{t}$ and the net variation per orbit $\ang{\Delta V}$ caused by a disturbing extra-acceleration. Our approach has a general validity since it is able to provide exact results for arbitrary orbital configurations. Furthermore, despite we focused only on some standard post-Keplerian dynamical features of motion such as general relativity to the first post-Newtonian level and the mass quadrupole moment, it can be extended to any perturbing effect of whatever physical origin, including, e.g., modified models of gravity. \textcolor{black}{As far as the Newtonian and post-Newtonian spin-dependent effects are concerned, our results hold for arbitrary orientations and sizes of the angular momenta and quadrupole moments of both the binary's bodies.}

We applied our results to the known stellar system surrounding the Supermassive Black Hole supposedly hosted by the Galactic Center at Sgr A$^\ast$.
In particular, the star S2, for which spectroscopic measurements of its radial velocity in the near-infrared accurate to about $\sigma_{V_\textrm{S2}}\simeq 30-50~\textrm{km~s}^{-1}$ are available since $t_0 = 2003.271$, will reach its periastron  next year after an orbital revolution 16 yr long. It could be a good opportunity to measure the static, Schwarzschild-like component of the post-Newtonian field of the Black Hole in Sgr A$^\ast$ which will change the stellar radial velocity by an amount as large as $\Delta V^\textrm{GE}_\textrm{max}\ton{t_\textrm{max}} = 551~\textrm{km~s}^{-1}$ at $t_\textrm{max} = 2018.35$. Instead, the resulting net shift per revolution, calculated for $t_0 = 2003.271$, will be as little as $\ang{\Delta V^\textrm{GE}} = -11.6~\textrm{km~s}^{-1}$. On the other hand, the periastron $\omega$ of S2 will be shifted by at most $0.2$ deg over the considered temporal interval, while the current uncertainty in estimating it from the observations amounts to $0.57$ deg\textcolor{black}{; the recent interferometer GRAVITY may improve such an accuracy level}. The other peculiar effects of the Kerr metric, i.e. frame-dragging and quadrupole mass moment, are far too small to be detected with S2 data; indeed, their maximum values, attainable for certain orientations of the Black Hole's spin axis, are as little as $0.19~\textrm{km~s}^{-1},0.0039~\textrm{km~s}^{-1}$, respectively. Our analytical results were confirmed also by numerical integrations of the equations of motion. Table \ref{resumeRV} resumes our findings.
Actually, it must be stressed that the impact of possible non-pointlike, diffused mass distribution in the Galactic Center due to, e.g., stellar remnants and/or Dark Matter on the radial velocity should be investigated as well. The same  holds also for the perturbations induced by other more or less distant individual stars inside or outside the S2's orbit which, however, should exhibit different temporal patterns with respect to the 1pN one of interest.
\textcolor{black}{Last but not least, also several effects pertaining the propagation of the electromagnetic waves contribute the spectroscopic measurements of S2 in such a way that they must be included in an overall analysis of the actual measurability of relativistic gravity in Sgr A$^\ast$; they should be the topic of dedicated analyses.}
\section*{Acknowledgements}
I am grateful to an attentive referee for her/his valuable and constructive comments which contributed to improve the manuscript.
\appendix
\section{Notations and definitions}\lb{appen}
Here, basic notations and definitions used in the text are presented \citep{1991ercm.book.....B,Nobilibook87,1989racm.book.....S,2003ASSL..293.....B}
\begin{description}
\item[] $G:$ Newtonian constant of gravitation
\item[] $c:$ speed of light in vacuum
\item[] A: binary's visible component
\item[] B: binary's invisible component
\item[] $\mA$: mass of the detectable body A
\item[] $\mB$: mass of the unseen companion B
\item[] $M_\textrm{tot}\doteq \mA + \mB$: total mass of the binary
\item[] $\mu\doteq GM_\textrm{tot}:$ gravitational parameter of the binary
\item[] $\xi\doteq\mA\mB M_\textrm{tot}^{-2}:$ dimensionless mass parameter of the binary
\item[] $S:$ magnitude of the angular momentum of any of the binary's components
\item[] $\mathcal{S}^\textrm{A/B}\doteq\ton{1+\rp{3}{4}\rp{M_\textrm{B/A}}{M_\textrm{A/B}}}S^\textrm{A/B}:$ magnitude of the scaled angular momentum of any of the binary's component
\item[] $\bds{\hat{S}}:$ unit vector of the spin axis of any of the binary's components
\item[] $\bds{\mathcal{S}}\doteq \sA+\sB:$ sum of the scaled angular momenta of the binary
\item[] $\chi_g:$ dimensionless angular momentum parameter of a Kerr black hole
\item[] $R:$ equatorial radius of any of the binary's components
\item[] $J_2:$ dimensionless quadrupole mass moment of any of the binary's components
\item[] $Q_2:$ dimensional quadrupole mass moment of any of the binary's components
\item[] ${\mathcal{Q}}_2^\textrm{A/B}\doteq\ton{1+\rp{M_\textrm{B/A}}{M_\textrm{A/B}}}Q_2^\textrm{A/B}:$ scaled dimensional quadrupole mass moment of any of the binary's components
\item[] $\mathbf{r}:$ relative position vector of the binary's orbit
\item[] $\mathbf{v}:$ relative velocity vector of the binary's orbit
\item[] $a:$  semimajor axis of the binary's relative orbit
\item[] $\nk \doteq \sqrt{\mu a^{-3}}:$   Keplerian mean motion
\item[] $\Pb = 2\uppi \nk^{-1}:$ Keplerian orbital period
\item[] $a_\textrm{A}=\mB M^{-1}_\textrm{tot} a:$ semimajor axis of the barycentric orbit of the binary's visible component A
\item[] $e:$  eccentricity
\item[] $p\doteq a(1-e^2):$  semilatus rectum
\item[] $I:$  inclination of the orbital plane
\item[] $\Omega:$  longitude of the ascending node
\item[] $\omega:$  argument of pericenter
\item[] $\varpi\doteq \Omega+\omega:$ longitude of pericenter
\item[] $t_p:$ time of periastron passage
\item[] $t_0:$ reference epoch
\item[] $\mathcal{M}\doteq \nk\ton{t - t_p}:$ mean anomaly
\item[] $\eta\doteq\nk\ton{t_0-t_p}:$ mean anomaly at epoch
\item[] $\lambda\doteq \varpi + \mathcal{M}:$ mean longitude
\item[] $\epsilon:$ mean longitude at epoch
\item[] $f:$  true anomaly
\item[] $f_0:$  true anomaly at epoch
\item[] $E:$ eccentric anomaly
\item[] $u\doteq \omega + f:$  argument of latitude
\item[] $\bds{\hat{l}}\doteq\grf{\cO,~\sO,~0}:$ unit vector directed along the line of the nodes toward the ascending node
\item[] $\bds{\hat{m}}\doteq\grf{-\cI\sO,~\cI\cO,~\sI}:$ unit vector directed transversely to the line of the nodes in the orbital plane
\item[] $r:$ magnitude of the binary's relative position vector
\item[] $\uR\doteq \mathbf{r}~r^{-1}=\bds{\hat{l}}\cos u + \bds{\hat{m}}\sin u:$ radial unit vector
\item[] $\uN\doteq\grf{\sI\sO,~-\sI\cO,~ \cI}:$ unit vector of the orbital angular momentum
\item[] $\uT\doteq\uN\bds\times\uR:$ transverse unit vector
\item[] $\bds A:$ disturbing acceleration
\item[] $A_{\rho}=\bds A\bds\cdot\uR:$ radial component of $\bds A$
\item[] $A_{\tau}=\bds A\bds\cdot\uT:$ transverse component of $\bds A$
\item[] $A_{\nu}=\bds A\bds\cdot\uN:$ normal component of $\bds A$
\item[] $V_\textrm{A}:$ radial velocity of the visible binary's component A
\end{description}
\section{Tables and Figures}
\begin{table*}
\caption{Relevant physical and orbital parameters of the star S2 and the SMBH at the GC along with their estimated uncertainties according to table 3 of \citet{2017ApJ...837...30G}; they are referred to the epoch $2000.0$. $D_0$ is the distance to $\textrm{Sgr~A}^\ast$. The Schwarzschild radius of the SMBH is $r_g\doteq 2GM_\bullet/c^2 = 0.088~\textrm{au}$, while the linear size of the semimajor axis of S2 is $a=1,044~\textrm{au}=11,863.6~r_g$. In footnote 4, pag. 2, \citet{2017ApJ...837...30G} write that their Keplerian orbital element $\omega$ would represent the longitude of periastron which, instead, is usually defined in the literature as the sum of the argument of periastron $\omega$ and the longitude of the ascending node $\Omega$, being dubbed as $\varpi$. Nonetheless, the numerical values quoted by \citet{2017ApJ...837...30G} in their table 3 for $\omega$ of various S stars turn out to correspond just to those for the argument of periastron, which enters our analytical formulas. We quote also the derived values of the SMBH's angular momentum and quadrupole mass moment calculated as \citep{1970JMP....11.2580G,1974JMP....15...46H} $S_\bullet=\chi_g M_\bullet ^2 G c^{-1},~Q_2^\bullet = - S^2_\bullet c^{-2} M^{-1}_\bullet$ due to the the \virg{no-hair} theorems \citep{1972CMaPh..25..152H,1967PhRv..164.1776I,1975PhRvL..34..905R}. The dimensionless parameter $\chi_g\leq 1$ is of the order of about $0.6$ for the SMBH in Sgr A$^\ast$ \citep{2016ApJ...818..121P}. The mass of the star S2 amounts to $M_\textrm{S2}=20~\textrm{M}_\odot$ \citep{2008ApJ...672L.119M}. We display also the value $f_0$ inferred from \rfr{fMt} for the true anomaly at the epoch $t_0=2003.271$, corresponding to the beginning of the radial velocity measurements used in table 5 of \citet{2017ApJ...837...30G}.}
\label{S2}
\centering
\begin{tabular}{ll}
\noalign{\smallskip}
\hline
Estimated parameter & Value \\
\hline
$M_\bullet$ & $4.28\pm \left.0.10\right|_\textrm{stat}\pm \left.0.21\right|_\textrm{sys}\times 10^6~\textrm{M}_\odot$\\
$D_0$ & $8.32\pm\left.0.07\right|_\textrm{stat}\pm \left.0.14\right|_\textrm{sys}~\textrm{kpc}$\\
$\Pb$ & $16.00\pm 0.02~\textrm{yr}$\\
$a$ & $0.1255\pm 0.0009~\textrm{arcsec}$\\
$e$  & $0.8839\pm 0.0019$\\
$I$ & $134.18\pm 0.40~\textrm{deg}$\\
$\Omega$ & $226.94\pm 0.60~\textrm{deg}$\\
$\omega$ & $65.51\pm 0.57~\textrm{deg}$\\
$t_p$ & $2002.33\pm 0.01$~\textrm{calendar~year}\\
\hline
Derived parameter & Value \\
\hline
$f_0$ & $139.72\pm 0.48$~\textrm{deg}\\
$S_\bullet$ & $\chi_g~1.61\times 10^{55}~\textrm{kg~m}^2~\textrm{s}^{-1}$ \\
$Q_2^\bullet$ & $-\chi_g^2~3.40\times 10^{56}~\textrm{kg~m}^2$\\
\hline
\end{tabular}
\end{table*}
\begin{table*}
\caption{Maximum and minimum \textcolor{black}{nominal} values of the pK radial velocity shifts of the star S2 along with the corresponding epochs and orientations of the SMBH'S spin axis calculated with the \textcolor{black}{central values} of Table \ref{S2}. The analytical expressions for $\Delta V^\textrm{GE}\ton{f},~\Delta V^\textrm{LT}\ton{f},~\Delta V^{Q_2}\ton{f}$, not displayed explicitly in the text because of their cumbersomeness, were used along with \rfr{fMt} to infer the epochs. Cfr. with the analytical and numerically integrated curves of Figures \ref{figura1} to \ref{figura3}. The published uncertainties in the measured values of the S2's RV are of the order of $\sigma_{V_\textrm{S2}}\simeq 30-50~\textrm{km~s}^{-1}$, as per table 5 of \citet{2017ApJ...837...30G}.}
\label{resumeRV}
\centering
\begin{tabular}{llll}
\noalign{\smallskip}
\hline
$\Delta V^\textrm{GE}_\textrm{max} = 551~\textrm{km~s}^{-1}$ & $t_\textrm{max}=2018.35$ & $-$ & $-$ \\
$\Delta V^\textrm{GE}_\textrm{min} = -70~\textrm{km~s}^{-1}$ & $t_\textrm{max}=2018.13$ & $-$ & $-$ \\
$\Delta V^\textrm{LT}_\textrm{max} = 0.19~\textrm{km~s}^{-1}$ & $t_\textrm{max}=2018.40$ & $i^\bullet_\textrm{max}= 160~\textrm{deg}$ & $\varepsilon^\bullet_\textrm{max}= 123 ~\textrm{deg}$\\
$\Delta V^\textrm{LT}_\textrm{min}= -0.18~\textrm{km~s}^{-1}$ & $t_\textrm{min}=2018.41$ & $i^\bullet_\textrm{min}= 12.3~\textrm{deg}$ & $\varepsilon^\bullet_\textrm{min}= 0~\textrm{deg}$\\
$\Delta V^{Q_2}_\textrm{max} = 0.0039~\textrm{km~s}^{-1}$ & $t_\textrm{max}=2018.348$ & $i^\bullet_\textrm{max}= 72.9~\textrm{deg}$ & $\varepsilon^\bullet_\textrm{max}= 207~\textrm{deg}$\\
$\Delta V^{Q_2}_\textrm{min}= -0.0022~\textrm{km~s}^{-1}$ & $t_\textrm{min}=2018.354$ & $i^\bullet_\textrm{min}= 145.4~\textrm{deg}$ & $\varepsilon^\bullet_\textrm{min}= 143.6~\textrm{deg}$\\
\hline
\end{tabular}
\end{table*}
%
%
%
%
%
%
%
%
%
%
%
%
%
%
%
%
%
%
%
%
%
%
\bibliography{PXbib,IorioFupeng}{}

\end{document}